\documentclass{ieeeaccess}
\usepackage{cite}
\usepackage[english]{babel}
\usepackage{algorithm}
\usepackage[utf8]{inputenc}
\usepackage[noend]{algpseudocode}
\usepackage{amsmath,amssymb,amsfonts}
\usepackage{verbatim}
\usepackage{graphicx}
\usepackage{textcomp}
\usepackage{soul,xcolor}
\usepackage{comment}
\usepackage{caption}
\usepackage{multirow}
\usepackage{color, colortbl}
\usepackage{subfigure}
\usepackage{hyperref}
\usepackage{nomencl}
\usepackage{wasysym}
\usepackage{soul}

\begin{document}
\history{Date of publication xxxx 00, 0000, date of current version xxxx 00, 0000.}
\doi{10.1109/ACCESS.2017.DOI}

\title{Deep Reinforcement Learning for Cybersecurity Assessment of Wind Integrated Power Systems}

\author{\uppercase{
\uppercase{Xiaorui Liu}\authorrefmark{1}, \IEEEmembership{Student Member, IEEE}, \uppercase{Juan Ospina}\authorrefmark{1}, \IEEEmembership{Member, IEEE} and \\ \uppercase{Charalambos Konstantinou}\authorrefmark{1}, \IEEEmembership{Senior~Member, IEEE}}
\address[1]{FAMU-FSU College of Engineering, Center for Advanced Power Systems, Florida State University}}

\tfootnote{This work was supported in part by Cyber Florida under Award \#3910-1011-00-A. }

\markboth
{Liu \headeretal: Deep Reinforcement Learning for Cybersecurity Assessment of Wind Integrated Power Systems}
{Liu \headeretal: Deep Reinforcement Learning for Cybersecurity Assessment of Wind Integrated Power Systems}

\corresp{Corresponding author: Charalambos Konstantinou (e-mail: ckonstantinou@ieee.org).}

\begin{abstract}
The integration of renewable energy sources (RES) is rapidly increasing in electric power systems (EPS). While the inclusion of intermittent RES coupled with the wide-scale deployment of communication and sensing devices is important towards a fully smart grid, it has also expanded the cyber-threat landscape, effectively making power systems vulnerable to cyberattacks. This paper proposes a cybersecurity assessment approach designed to assess the cyberphysical security of EPS. The work takes into consideration the intermittent generation of RES, vulnerabilities introduced by microprocessor-based electronic information and operational technology (IT/OT) devices, and contingency analysis results. The proposed approach utilizes deep reinforcement learning (DRL) and an adapted Common Vulnerability Scoring System (CVSS) score tailored to assess vulnerabilities in EPS in order to identify the optimal attack transition policy based on $N-2$ contingency results, i.e., the simultaneous failure of two system elements. The effectiveness of the work is validated via numerical and real-time simulation experiments performed on literature-based power grid test cases. The results demonstrate how the proposed method based on deep $Q$-network (D$Q$N) performs closely to a graph-search approach in terms of the number of transitions needed to find the optimal attack policy,  without the need for full observation of the system. In addition, the experiments present the method’s scalability by showcasing the number of transitions needed to find the optimal attack transition policy in a large system such as the Polish 2383 bus test system. The results exhibit how the proposed approach requires one order of magnitude fewer transitions when compared to a \emph{random} transition policy.

\end{abstract}

\begin{keywords}
Cybersecurity assessment, contingency analysis, cyberattacks, deep reinforcement learning.
\end{keywords}

\titlepgskip=-15pt
\maketitle

\section{Introduction}

The power grid is the cornerstone of all critical infrastructures. The safe and secure functionality of electric power systems (EPS) is directly related to every aspect of the economy and society. In the last decades, worldwide energy demand has significantly increased and is estimated to continue to do so by nearly 50\% by 2050 \cite{eia2019_energydemand}. Due to the increasing energy demand as well as the need to enhance system efficiency and asset reliability, the technological modernization of the power grid infrastructure has become an immediate priority for governments and energy stakeholders around the world \cite{muyeen2017communication}. This modernization, alongside environmental concerns, are driving factors for the integration of renewable energy sources (RES) to the power grid. For example, the U.S. Energy Information Administration (EIA) indicates that, in 2019, the wind was responsible for generating approximately 42\% of RES generated power at utility-scale facilities in the U.S., and 7.3\% of the total U.S. electricity generation, making it the most popular RES \cite{RESWIND}. Even though wind integration aids in accommodating the increasing power demand, its intermittent nature introduces challenges related to the mismatch between supply and demand. For instance, short-term wind power fluctuations occur on a second or sub-second timescale during which load balancing methods do not yet operate. Thus, to ensure system stability, critical aspects such as optimal location, power flow, and generation variance must be taken into consideration when interconnecting wind energy systems. 

Traditionally, contingency analysis has been used to assess physical power system security in EPS \cite{balu1992line}. This is achieved by calculating the power flow of all the system elements in the event of single or multiple failures. In essence, a contingency is the failure or loss of any element such as a circuit breaker, generator, or transmission line. Contingencies can be planned or unplanned. Planned contingencies include events resulting from scheduled maintenance and proactive emergency preparedness, while unplanned events include fluctuating wind injections, cyberattacks, human errors, etc. The North American Electric Reliability Corporation (NERC) requires system operators to meet the $N-1$ security constraint and classifies systems into four main categories \cite{chen2013n}. These categories are shown in Table \ref{table:contingencycats}.

\begin{table}[t]
\centering
\caption{NERC TPL-001-4 contingency categories.}
\label{table:contingencycats}
\begin{tabular}{||c|c||}
\hline\hline
\textbf{Category} & \textbf{Contingency Case} \\ \hline
$A$ & No contingency \\ \hline
$B$ & $N-1$ \\ \hline
$C$ & $N-1-1$ or $N-2$ \\ \hline
$D$ & $N-k$, $k>2$ (cascading) \\ \hline\hline
\end{tabular}
\end{table}

The intermittent nature of wind power generates challenges when performing security studies of power systems based on contingency analysis. Their intermittency can rapidly change the most critical contingencies of the system or create a number of contingencies ($\lambda$) that exceeds the maximum number of contingencies that the system can handle ($k$); thus leading to cascading scenarios. A prime example of insufficient security margins is the widespread power outage across the U.K. in 2019 \cite{UK}. The near-simultaneous loss of two-generation sites, one being an offshore wind farm and the other one a gas-fired power station, resulted in a massive under-frequency event. Load shedding mechanisms responded immediately causing a major disturbance that affected nearly one million people.

Additionally, the power grid is experiencing a rapid move towards a more interconnected system. Operational technology (OT) electronic devices are deployed and operated at all scales of the power system and are often being designed and retrofitted with information technology (IT) devices to support communication processes and protocols that enhance the controllability and observability of the system. The use of such digital electronic devices with software applications \cite{stright2020defensive}, modules, drivers, commercial-off-the-shelf (COTS) hardware \cite{konstantinou2019hardware}, and network resources is a double-edged sword \cite{ospina2020trustworthy}. On one hand, it assists in the development of the future modern and advanced grid in terms of optimizing asset utilization, addressing disturbances, providing better power quality, and accommodating all storage and generation options with grid-support functions. On the other hand, the coupling between such cyber-electronic devices and physical components in power systems has altered the threat model \cite{iet:/content/books/10.1049/pbpo095e_ch15, mclaughlin2016cybersecurity}. 

In the past, the threat model has been solely focused on physical threats. However, due to the integration of such network-controlled components, the security challenges need to consider both the cyber and physical nature of the grid, addressing the growing number of emerging threats \cite{gritzalis2019critical}. Some examples of these potential threats are presented in \cite{keliris2018low,keliris2019open,liu2018assessment}, where it has been demonstrated that attackers can leverage publicly available sources by using open-source intelligence (OSINT) techniques combined with open-source exploitation methods in order to spoof GPS signals of phasor measurement units (PMUs). Another example is presented in \cite{case2016}, where a real-world attack within the Ukrainian power system is accomplished by injecting malicious firmware in serial-to-Ethernet gateways at targeted substations. Attackers were able to trip circuit breakers and cause a blackout that affected approximately 225,000 customers.

In this work, we provide an effective way for system operators, at both the local and international level, to assess the cyberphysical security of EPS, which takes into consideration wind uncertainty together with cyber-based aspects such as quantitative scoring systems of vulnerabilities identified in IT/OT devices supporting the grid infrastructure. The assessment follows a step-by-step process, from an attacker's point of view, designed to identify the most critical system points an adversary can leverage to compromise the targeted EPS. Our contributions are summarized as follows:

(1) We propose a cybersecurity assessment approach that considers adversaries that make use of OSINT modeling techniques to construct power system models. Such models are then used in tandem with contingency analysis that takes into account wind intermittent generation to identify the critical cyber and physical vulnerabilities of the EPS. The assessment process is performed without the need for full observability of the system since it models the state of the power system as a partially observable Markov decision process (POMDP) that is solved using deep $Q$-networks (D$Q$Ns). The solution given by the proposed D$Q$N reveals the optimal attack transition policy an adversary would follow to potentially induce cascading failures in the assessed cyberphysical EPS.
	
(2) We propose an adapted version of the Common Vulnerability Scoring System (CVSS) based on contingency analysis results and information from the power and communication networks that reveal cyber and physical vulnerabilities within system nodes. The adapted CVSS is used to generate a transition graph designed to assess the complexity of each possible attack path based on various adversarial strategies.
		
(3) We evaluate the performance of the proposed methodology using real-time simulations on test power systems highlighting the method’s scalability. Our results showcase that a fewer number of transitions is needed compared to a \emph{random} transition policy in order to find the optimal attack transition.

The rest of the paper is organized as follows. Section \ref{s:Rela} presents related work. In Section \ref{s:method}, we introduce the methodology of the proposed cybersecurity assessment approach. Section \ref{s:result} presents contingency analysis studies for various power system test cases using real-time simulation. In Section \ref{s:result2}, we demonstrate the effectiveness of the proposed cybersecurity assessment approach and compare it to different transition techniques. Finally, Section \ref{s:conclusion} concludes the paper and provides directions for future work.

\section{Related Work}\label{s:Rela}
In this part, we explore some of the state-of-the-art approaches being proposed by researchers that aim to address issues related to (1) $N-k$ contingencies simulations considering intermittent RES generation, (2) assessing the severity of electronic devices security vulnerabilities, and (3) vulnerability and risk assessments methods for cyberphysical EPS.

\subsubsection{Methods for $N-k$ Contingency Analysis}
Towards reducing the occurrence of cascading failures, existing research efforts have focused on proposing efficient methods that can perform studies based on $N-k$ contingency scenarios. Due to the size and complexity of power systems, these ``what-if'' contingency scenarios are based on computationally expensive optimal power flow processes. Research in this area aims to address the computational overhead of $N!/[k!(N-k)!]$ simulations for $N-k$ contingencies. For example, the work in \cite{ejebe1988fast} describes a fast-bounding case which requires a small online memory model. Other efforts compute the active power flow change at lines and the voltage change at buses to evaluate the severity of $N-1$ and $N-2$ contingencies \cite{burada2016contingency}. In \cite{zhao2018graph}, a graph-based power model analysis is presented for contingency ranking. In \cite{hasan2017heuristics}, a heuristics pruning approach for identifying $N-k$ contingencies is discussed while a topology-based algorithm that considers whether the generator or line is in densely populated areas is presented in \cite{vellaithurai2014cpindex}. The authors use the concepts of closeness and betweenness centrality to determine the component's importance for a $N-k$ criterion. 

One of the main challenges of performing contingency studies in power systems with high penetration of wind is that the uncertain nature of wind causes high variability when identifying the most critical $N-k$ contingencies of the system. Existing studies do not often take into account this variability \cite{vrakopoulou2013probabilistic, sundar2016unit,bai2016robust}. 
The authors in \cite{sundar2016unit} and \cite{morales2010probabilistic} demonstrate some of the effects that intermittent power generation has in critical contingency identification. They present probabilistic power flow studies that show how the variable nature of power flow, due to wind fluctuations and uncertainties, can alter the number and location of the most critical contingencies recognized by system operators. The correct identification of these critical contingencies is of paramount importance as they can be potentially leveraged by adversaries in order to cause major disruptions in EPS  \cite{konstantinou2016attacking, kelirisbh17}.

\subsubsection{Severity Assessment of Electronic Devices Security Vulnerabilities}
The wide-scale integration of information and communication technologies in the form of digital electronic devices into the electrical grid expands the list of possible attack vectors that adversaries could exploit to cause major disruptive events. Hence, in order to ensure the secure operation of the entire system, it is essential to consider the inherent vulnerabilities introduced by the grid-supporting IT/OT infrastructure. One scoring system that is widely used for device-level vulnerability assessments in the IT industry is CVSS \cite{cvssreport}. The CVSS can assess the severity of software, hardware, and firmware vulnerabilities by using numerical scores. One example of its use can be found in \cite{zhang2015power}. Here, the authors utilize CVSS to estimate the probability of successfully exploiting identified independent vulnerabilities, including zero-days, existing in components connected to the LAN of a supervisory control and data acquisition (SCADA) system. Another example of CVSS use can be found in \cite{venkataramanan2019cyphyr}, where a CVSS-based cyber asset impact score is presented providing a real-time cyber impact severity score that can be used as a basis for processes such as vulnerability management, isolation of cyber assets, and system reconfiguration.

\subsubsection{Vulnerability and Risk Assessment Methods for Cyberphysical EPS}
Several researchers have focused on developing system-wide security assessment tools aimed to identify possible vulnerabilities and attack vectors which can be subsequently used to produce optimal control policies designed to guide secure operations of cyberphysical EPS. The work presented in \cite{huang2019adaptive}, proposes power system emergency control mechanisms based on D$Q$Ns to maintain the reliable operation of the system by performing dynamic breaking of generation and under-voltage load shedding. Other researchers have also made use of D$Q$Ns to perform cybersecurity analysis studies in EPS. One example is presented in \cite{wang2020coordinated}, where the authors propose a D$Q$N-based cyberphysical topology attack designed to trip critically-targeted transmission lines with the objective of exceeding power flow line constraints in the system. This research demonstrates how the disconnection of essential transmission lines could cause system collapse, and how attackers can find out what type of topology attack needs to be performed in order to cause cascading failures. A similar approach is taken by researchers in \cite{9087633}, where a robust D$Q$N-based contingency management mechanism is proposed to provide remedial action when contingencies exist in the system.

\begin{figure*}[ht]
\centerline{\includegraphics[width=0.9\textwidth]{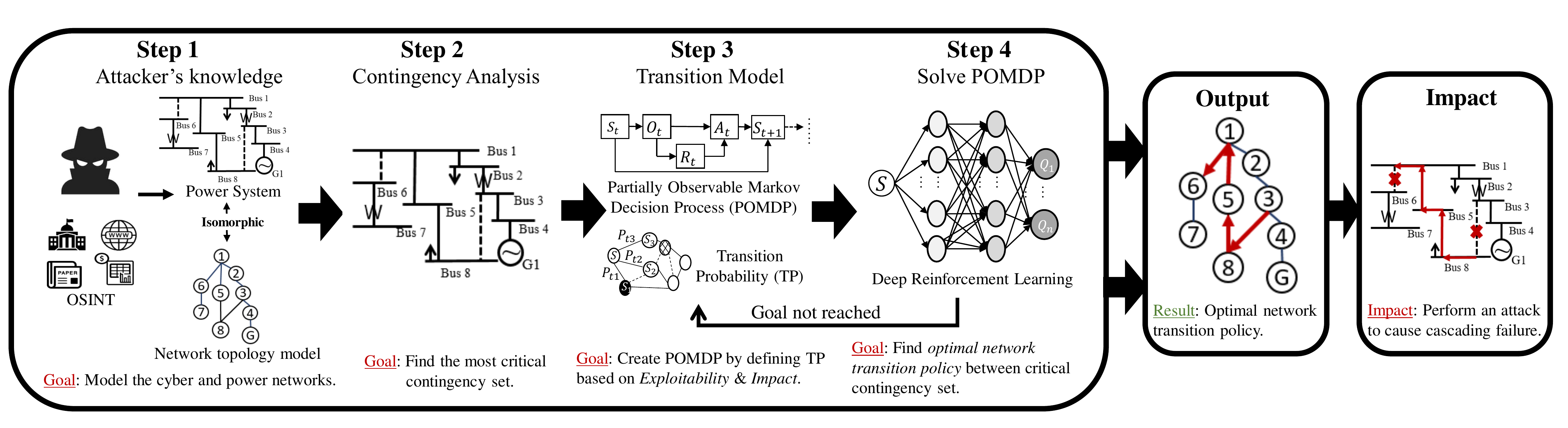}}
\caption{Graphical depiction of the major steps of the proposed cybersecurity assessment process and the optimal attack transition policy given as output.}
\label{fig:architecture}
\end{figure*}

Other works have focused on more traditional ranking mechanisms to improve EPS cybersecurity. For example, the research presented in \cite{vellaithurai2014cpindex} assesses system vulnerability from the cyberphysical security perspective using contingency ranking methods and a cyber-intrusion ranking methodology. Similarly, in \cite{davis2015cyber}, an operational reliability impact assessment framework has been developed. In this study, the authors incorporate cyberphysical threats in the assessment of the EPS  operation. Another approach is presented in \cite{mehdizadeh2018power}, describing an overload risk assessment method based on $N-1$ contingency analysis and wind penetration.

Compared to the existing work, our proposed cybersecurity assessment approach assesses the cyberphysical security of EPS leveraging the use of deep reinforcement learning (DRL) paradigms while considering the intrinsic interactions between the system’s network and physical components. Our work considers physical-based aspects, such as contingency analysis and wind uncertainty, together with network-based aspects, such as vulnerabilities stemming from OSINT methods. The proposed assessment framework reveals potential threats that can be utilized by attackers to cause serious disruptions in EPS, and at the same time, assist system operators to prioritize the deployment of cybersecurity mechanisms, and thus, contribute towards reducing the risk of cyberattacks.

\section{Cybersecurity Assessment Methodology}\label{s:method}

In this section, we provide the methodology of the proposed approach. Fig. \ref{fig:architecture} shows the step-by-step process that our cybersecurity assessment approach follows.  In step 1, the assessment process determines the threat model based on the adversary objectives and capabilities. Specifically, our work considers an attacker that leverages OSINT techniques. The OSINT methods are utilized in step 2 in order to run contingency analysis with the objective of identifying the set of \textit{k} critical contingencies of the system. Our results focus on two contingencies that assess the power system condition when two components are lost, i.e., $k = 2$. However, the proposed approach can be extended to consider a higher number of contingencies. To proceed with the assessment process without the need for full system observability \cite{8810568}, in step 3, the proposed approach creates a POMDP by defining a transition probability ($TP$) based on the proposed adapted version of the CVSS score metric. The score evaluates the difficulty of each network transition in the generated system graph. Then, in step 4, the POMDP is solved using a DRL model designed to find the optimal attack policy between the previously identified contingencies. Finally, the output of the cybersecurity assessment process evaluates the potential threat by revealing the optimal attack transition policy between the identified contingency pair which could cause cascading failures in the physical system. The details of each step are presented in the following subsections. 

\subsection{Step 1: Threat Model}
OSINT refers to a collection of techniques and methodologies that can be used to gather, analyze, and exploit publicly available information (e.g., public government data, commercial data,  social media, etc.) to characterize aspects of a particular target system. Existing work has demonstrated that the U.S. power grid infrastructure could be effectively profiled using OSINT techniques \cite{OSINT, ics}. Following a similar approach, we consider a threat model where an attacker can leverage publicly available information using OSINT methods to collect sufficient EPS data (e.g., line parameters, the status and location of circuit breakers, system topology, generation sites and capacity, etc.). Also, the attacker is able to acquire data to calculate power flow and therefore run contingency analysis \cite{keliris2019open}. Depending on the degree of system contingencies (e.g., $N-1$ secure system), the adversary can leverage the ranking results of contingency algorithms to identify which system elements if  ``removed'' can lead to an insecure power system state. Although a plethora of public power system information is available, it is unlikely that the attacker will ever have full knowledge and real-time observability of the system \cite{morere2016bayesian}. In our approach, it is assumed that the attacker, in spite of having the necessary information to perform contingency analysis via OSINT techniques, he/she does not have the full state information of the system. Specifically, while the adversarial agent is transitioning through the cyber system network to exploit vulnerabilities in the identified double contingency nodes, he/she is unaware of his/her position relative to the contingencies and the cyber network transition complexities (based on the adapted CVSS) of the different attack paths.

In addition, we assume that \emph{the cyber system network graph is isomorphic with the physical system graph}, indicating that the topology of the communication network is mapped with the topology of the physical system. Therefore, we model the environment as a POMDP in which the agent may only access the current state and make an observation for obtaining possible actions in each state (Step 3). Based on the observation results for each state-action combination, the network $TP$ is calculated. This probability reveals the transition complexity between different states. By leveraging this methodology, a D$Q$N-based algorithm is then utilized to identify the optimal attack transition policy between the critical contingency elements (Step 4). 

\subsection{Step 2: Contingency Analysis}\label{ss:contingencystep2}

In order to find the attacker's optimal attack transition policy, we first need to identify the set of critical double contingencies of the physical power system (e.g., simultaneous $N-2$ or consecutively $N-1-1$). We utilize a fast pruning $N-2$ algorithm to find all the thermal constraint violations via linear power flow approximation \cite{turitsyn2013}. The algorithm is initiated based on the set of all $N-2$ pairs. The contingency candidate list is pruned using line outage distribution factors (LODFs). LODFs describe the power flow impact on other lines when a line outage occurs. The pruning approach is based on the thermal constraints of lines, running until the number of contingencies included in the set does not change. If the LODF exceeds its thermal constraint, it is added to the contingency candidate set. The line overload condition can be written as $A_{xy}\cdot B_{xc}+A_{yx}\cdot B_{yc} > 1$, where \textit{x} and \textit{y} are lines experiencing outages, \textit{z} is an arbitrary line experiencing power flow changes, and \textit{c} is a possible constraint. Matrix $A_{xy}$ can be calculated by $A_{xy}=(1+L_{xy}\cdot f_y/f_x)/(1-L_{yx}\cdot L_{xy})$, where $L$ is the LODF shown in Eq. (\ref{eq:SingleLODF}). This equation describes the change in the flow through line $x$, where $f_x$ is the original flow, and $f'_x$ is the flow after the outage. Correspondingly, $f_y$ represents the flow through line $y$ before the line is tripped. Matrix $B_{xc}$ is calculated by $B_{xc}=f_x\cdot L_{zx}/(f_z^{critical}\pm f_z)$, where $f_z^{critical} $ is the bound value, and the $\pm$ sign corresponds to the conditions $f'_z < - f_z^{critical} $ and $f'_z >  f_z^{critical} $, respectively. Eq. (\ref{eq:2LODF}) shows the power flow variance experienced by line $z$ when lines $x$ and $y$ are experiencing outages. 

\begin{equation}
L_{xy} = \frac{f_{x}^{'}-f_{x}}{f_{y}}
\label{eq:SingleLODF}
\end{equation}

\begin{equation}
f_{z}^{'}-f_{z} = \frac{L_{zx} \cdot (f_{x}+L_{xy}\cdot f_y)}{1-L_{yx}\cdot L_{xy}} + \frac{L_{zy} \cdot (f_{y}+L_{yx}\cdot f_x)}{1-L_{yx}\cdot L_{xy}}
\label{eq:2LODF}
\end{equation}

\subsection{Step 3: POMDP Transition Model Based on Adapted CVSS}

After finding the most critical contingency set, the process advances to create the corresponding POMDP of the cyberphysical-graph environment by calculating the corresponding $TP$ between the different nodes of the system. Generally, POMDPs are used to model the response and outcomes of systems when different actions are performed at specific states. In our environment, observations made by the attacker do not provide full state information, i.e., the agent does not know apriori how many nodes the system has nor their respective states, and he/she needs to observe the environment to determine potential actions, hence the selection of POMDP system modeling. POMDPs can be mathematically modeled as a $6$-tuple ${\displaystyle(\mathcal{S},\mathcal{A},{\Omega},P,R,\mathcal{O})}$, where $\mathcal{S}$ is the set of all possible states in a given environment, $\mathcal{A}$ contains all the agent's potential actions, {$\Omega$} is a set which includes all possible observations, $P$ is the $TP$ for each state, $R$ is the reward function for performing different actions, and $\mathcal{O}$ represents conditional observation probabilities. The notation of POMDP tuples is summarized in Table. \ref{table:POMDPparameters}. At the current state $s$, given the $TP$ and observation $o$, the agent takes action $a$ to move to the next state $s'$. As a result of this state-action pair, the agent receives reward $R$. This process repeats until the terminal state is reached. In this POMDP formulation, the $TP$ for each state is an essential factor that must be determined adequately according to the process being modeled. In our case, the $TP$ relies on the cyber system vulnerabilities, i.e., vulnerabilities that exist in electronic devices, and their potential impacts related to the physical system, i.e., the identified power system contingencies.

\begin{table}[]
\centering
\caption{Notation of POMDP tuples.}
\label{table:POMDPparameters}
{
\begin{tabular}{||c|c||}
\hline\hline
\multicolumn{2}{||c||}{\textbf{POMDP tuples}} \\ \hline\hline
$\mathcal{S}$ &State set\\ \hline
$\mathcal{A}$  &Action set\\ \hline
$\Omega$ & Observation set\\\hline
$P$ ($TP$) &  Transition probability\\ \hline
$R$ & Reward function \\ \hline
$O $& Observation probabilities set \\ \hline
\hline
\end{tabular}}
\end{table}

\begin{figure}[t]
\centerline{\includegraphics[width=0.45\textwidth]{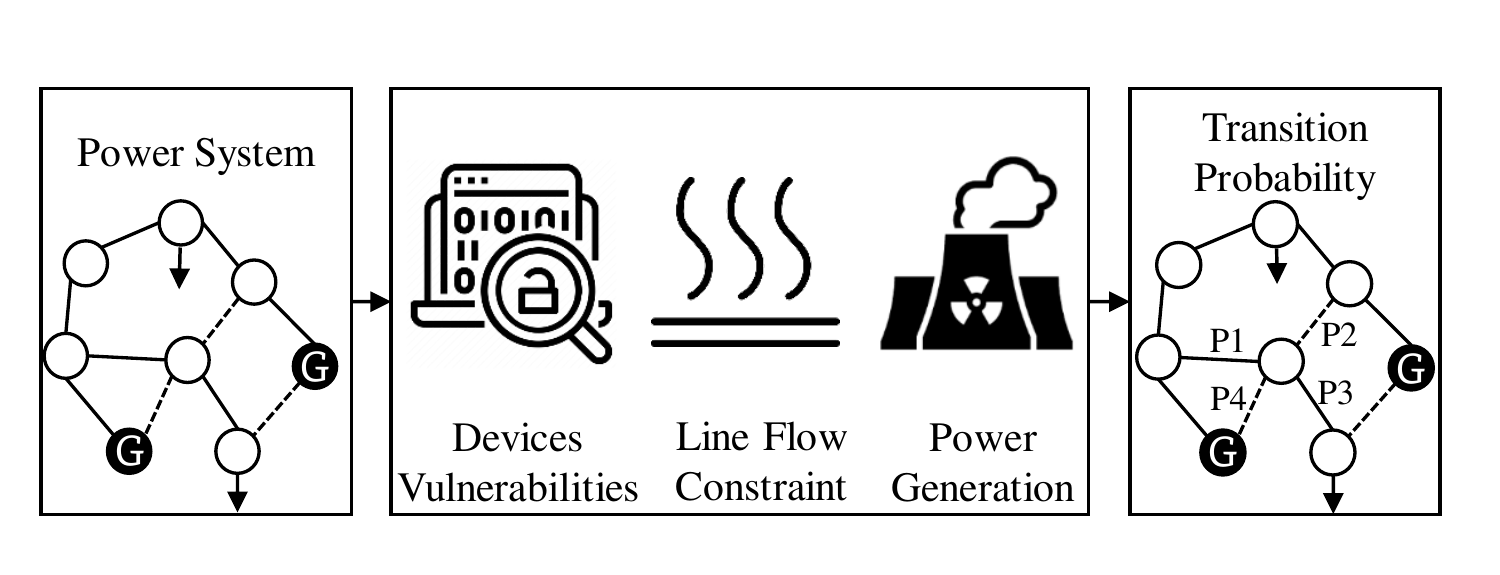}}
\caption{Overview of the transition probability ($TP$)  assessment.}
\label{fig:Trans}
\end{figure}

Considering the cyber network system vulnerabilities as well as the optimal attack transition policy between the identified contingencies (physical vulnerabilities), a $TP$ for each transition step (between cyberphysical system nodes) can be determined. These probabilities aid in the traversal agent's decision making since the values reveal the difference in complexity and difficulty for each transition, i.e., how vulnerable is the cyberphysical system at each node, i.e., bus, from the point of view of the attacker transition policy. In each step, the node's identified cyber and physical characteristics including the electronic device vulnerabilities, thermal limits of lines, and power generation are considered. A graphical illustration of this procedure is shown in Fig. \ref{fig:Trans}. In this work, we compute the $TP$ using an adapted version of CVSS v3.1. 

\begin{figure}[t]
\centerline{\includegraphics[width=0.35\textwidth]{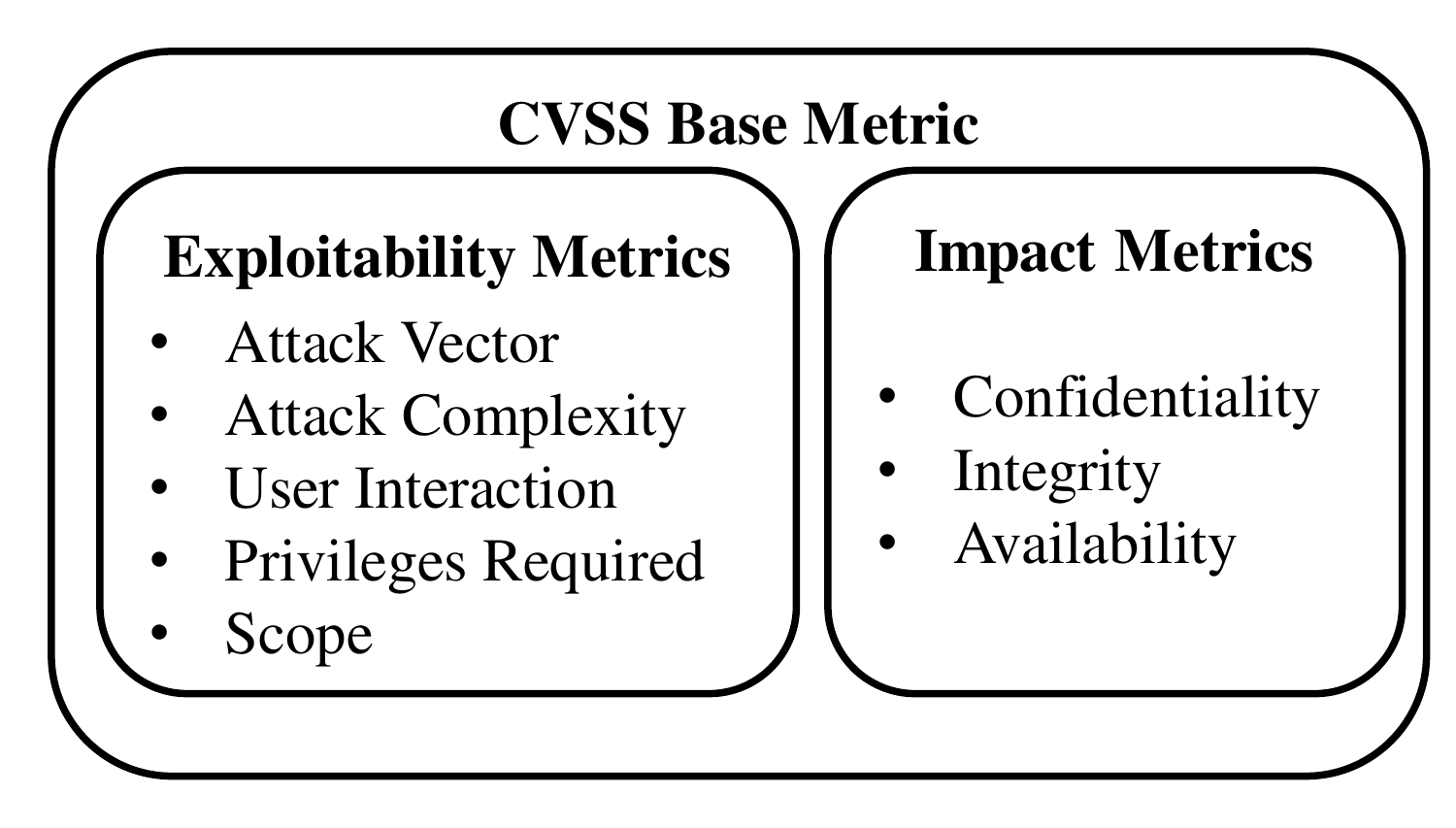}}
\caption{Outline of the Common Vulnerability Scoring System (CVSS) base metric.}
\label{fig:CVSS}
\end{figure}

CVSS is a vulnerability scoring system generally used in the IT industry to assess the severity of the identified computer system's vulnerabilities. Although there exist temporal and environmental metrics in CVSS, their main aim is to reflect how vulnerabilities change over time or demonstrate uniqueness to a particular user's environment \cite{cvssreport}. For our application, base metrics portray a better picture regarding how the cyber and physical vulnerabilities at each power system node affect the transition difficulty of the threat. More specifically, the base score provides a comprehensive assessment of the intrinsic characteristics of identified vulnerabilities using quantitative \textit{Exploitability} and \textit{Impact} metrics as shown in Fig. \ref{fig:CVSS}. The range of scores goes from $0$ to $10$, with $10$ being the most severe -- maximum value.

\subsubsection{Exploitability Metric}

This metric describes the difficulty and technical means by which software, hardware, or firmware vulnerability can be exploited. In our case, the exploitability represents the difficulty of vulnerability exploitation for each electronic device that exists in a particular node of the cyber-layer of the power system. In other words, it represents the complexity of the transition based on the type of node (i.e., $PQ$ or $PV$ power system bus) to which the agent is transitioning to. The overall score of this metric is determined by five sub-metrics, described below.

\textit{a) Attack Vector (AV) }--
This metric is defined as one of the following categories: \emph{network}, \emph{adjacent network}, \emph{local network}, or \emph{physical}. In a network attack, an adversary exploits a vulnerable device bound to the network stack. This type of attack is conducted through the Open Systems Interconnection (OSI) layer 3. In an EPS, an attacker may conduct a network attack by manipulating TCP-level packets flowing across a substation network. In an adjacent attack, the adversary also exploits vulnerable devices bound to the network stack. However, the attack cannot be performed across the boundary of OSI layer 3. In essence, the attack is limited to the same shared physical or logical network. An example of this type of attack is an Address Resolution Protocol (ARP) flooding attack that leads to a denial-of-service targeted at the control and monitoring devices connected to a LAN segment of a microgrid \cite{habib2017adaptive}. In a local attack, a direct path to the vulnerable element is required (e.g., local terminal, remote terminal, or deceive legitimate users into executing malicious instructions). In an EPS, this type of attack could be performed by executing malicious code in local control or monitoring electronic devices accessed via a local or remote terminal. Finally, for a physical attack, actual physical interaction between the attacker and the target is necessary. In an EPS, this means that the attacker must compromise the targeted electronic devices through physical means (e.g., causing physical damage to the devices).

\textit{b) Attack Complexity (AC)} -- 
This metric represents the amount of effort an attack on the vulnerable electronic device would require. The value of this metric, \emph{high} or \emph{low}, depends on the security level of the electronic devices as well as the adversary's capabilities and skills. In EPS, generation buses can be considered of more significance than load buses in regards to power grid operation and, consequently, possible threats. Typically, additional security mechanisms are in place to protect bulk generation infrastructure \cite{NERCsecurity}. This is accomplished by using electronic security devices, physical barriers, or security monitoring equipment. Hence, as part of our CVSS vulnerability scoring, $PV$ and $PQ$ buses are considered of \emph{high} and \emph{low} attack complexity, respectively.

\textit{c) User Interaction (UI)} -- 
This metric reveals whether user interaction is required to exploit a certain electronic device. It quantifies the amount of participation required from a human user, different from the attacker, to successfully compromise the targeted device. For example, attackers could attempt to deceive the system operator to give them access to the control room via phishing or malware attacks. Due to the importance of $PV$ buses, we assume that the attacker will require UI to manipulate a $PV$ bus. On the contrary, it is assumed that attackers would not need to obtain special permission from another human user to access $PQ$ buses. The values for this metric are: \emph{required} for $PV$ buses, and \emph{none} for $PQ$ buses.

\textit{d) Privileges Required (PR)} -- 
This metric determines the level of privileges needed to carry out an attack, i.e., it evaluates the level of privileges that are required by the attacker before successfully compromising the vulnerable electronic device. Similarly to the previous metric, we designate its values according to the type of power system bus being evaluated: \emph{high} for $PV$ buses,  and \emph{low} for $PQ$ buses.

\textit{e) Scope (S)} -- 
This metric demonstrates whether or not compromising a particular electronic device will cause implications beyond its security scope. If the scope metric is defined as \textit{changed}, attacking the corresponding electronic device will result in a detrimental implication beyond its security scope, i.e., will affect the other elements in the system. If the scope is defined as \textit{unchanged}, it will only cause implications to elements under the same security scope. In our context, when a $PQ$ bus is attacked, no major disturbances are observed in other system's elements since generation is not directly affected, thus its scope can be defined as \textit{unchanged}. However, if a $PV$ bus is compromised, more severe effects on surrounding nodes of the physical EPS network, caused by power stability issues, are observed. In this case, the scenario needs to be characterized by \textit{changed} scope.

Following the description of the exploitability metrics, Table \ref{table:scoresystems} shows a detailed comparison between the metrics values found in different available scoring systems. These scoring systems are the CVSS v3.1 \cite{cvssreport}, CVSS v2.0 \cite{cvss20_cite}, and the Industrial Vulnerability Scoring System (IVSS) \cite{ivss_cite}. CVSS v3.1 is the most up to date scoring system which provides the most accurate way of capturing the main characteristics of vulnerability via numerical scores. IVSS is an outdated scoring system and not widely used and supported by the community. Other quantitative risk assessment scoring systems, such as CCSS \cite{scarfone2010common} and CMSS \cite{van2009common}, were also considered when selecting the appropriate scoring system. However, all of these scoring systems are based on the previous version of CVSS, i.e., CVSS v2.0.

\begin{table}[]
\setlength{\tabcolsep}{1.8pt}
\centering
\caption{Exploitability submetrics comparison of different score systems: CVSS v3.1, CVSS v2.0, and IVSS.}
\label{table:scoresystems}
\begin{tabular}{||c|c|c|c|c||}
\hline\hline
\textbf{\begin{tabular}[c]{@{}c@{}}Score\\ System\end{tabular}} & \textbf{Metric} & \textbf{Abb.} & \textbf{\begin{tabular}[c]{@{}c@{}}Metric\\ Value\end{tabular}} & \textbf{\begin{tabular}[c]{@{}c@{}}Num.\\ Value\end{tabular}} \\ \hline \hline
\multirow{11}{*}{\textbf{\begin{tabular}[c]{@{}c@{}}CVSS \\ v3.1\end{tabular}}} & \multirow{4}{*}{\begin{tabular}[c]{@{}c@{}}Attack\\ Vector\end{tabular}} & \multirow{4}{*}{AV} & Network & 0.85 \\ \cline{4-5} 
 &  &  & \begin{tabular}[c]{@{}c@{}}Adjacent \\ network\end{tabular} & 0.62 \\ \cline{4-5} 
 &  &  & \begin{tabular}[c]{@{}c@{}}Local \\ network\end{tabular} & 0.55 \\ \cline{4-5} 
 &  &  & Physical & 0.2 \\ \cline{2-5} 
 & \multirow{2}{*}{\begin{tabular}[c]{@{}c@{}}Attack\\ Complexity\end{tabular}} & \multirow{2}{*}{AC} & Low & 0.77 \\ \cline{4-5} 
 &  &  & High & 0.44 \\ \cline{2-5} 
 & \multirow{2}{*}{\begin{tabular}[c]{@{}c@{}}User \\ Interaction\end{tabular}} & \multirow{2}{*}{UI} & None & 0.85 \\ \cline{4-5} 
 &  &  & Required & 0.62 \\ \cline{2-5} 
 & \multirow{3}{*}{\begin{tabular}[c]{@{}c@{}}\\ Privileges \\ Required\end{tabular}} & \multirow{3}{*}{PR} & None & 0.85 \\ \cline{4-5} 
 &  &  & Low & \begin{tabular}[c]{@{}c@{}}0.62 if S = Unchanged\\ 0.68 if S = Changed\end{tabular} \\ \cline{4-5} 
 &  &  & High & \begin{tabular}[c]{@{}c@{}}0.27 if S = Unchanged\\ 0.50 if S = Changed\end{tabular} \\ \hline \hline
\multirow{9}{*}{\textbf{\begin{tabular}[c]{@{}c@{}}CVSS \\ v2.0\end{tabular}}} & \multirow{3}{*}{\begin{tabular}[c]{@{}c@{}}\\  Access (Attack)\\ Vector\end{tabular}} & \multirow{3}{*}{AV} & Network & 1.0 \\ \cline{4-5} 
 &  &  & \begin{tabular}[c]{@{}c@{}}Adjacent \\ network\end{tabular} & 0.646 \\ \cline{4-5} 
 &  &  & \begin{tabular}[c]{@{}c@{}}Local \\ network\end{tabular} & 0.395 \\ \cline{2-5} 
 & \multirow{3}{*}{\begin{tabular}[c]{@{}c@{}}Access (Attack)\\ Complexity\end{tabular}} & \multirow{3}{*}{AC} & Low & 0.71 \\ \cline{4-5} 
 &  &  & Medium & 0.61 \\ \cline{4-5} 
 &  &  & High & 0.35 \\ \cline{2-5} 
 & \multirow{3}{*}{\begin{tabular}[c]{@{}c@{}}Authentication\\ (v3.1 - Privileges \\ Required)\end{tabular}} & \multirow{3}{*}{\begin{tabular}[c]{@{}c@{}}AU\\ (PR)\end{tabular}} & None & 0.704 \\ \cline{4-5} 
 &  &  & Single & 0.56 \\ \cline{4-5} 
 &  &  & Multiple & 0.45 \\ \hline \hline
\multirow{12}{*}{\textbf{IVSS}} & \multirow{4}{*}{\begin{tabular}[c]{@{}c@{}}\\ Access (Attack)\\ Vector\end{tabular}} & \multirow{4}{*}{AV} & \begin{tabular}[c]{@{}c@{}}Remote \\ (v3.1 Network)\end{tabular} & 1.0 \\ \cline{4-5} 
 &  &  & \begin{tabular}[c]{@{}c@{}}Local network \\ (v3.1 Adjacent)\end{tabular} & 0.7 \\ \cline{4-5} 
 &  &  & \begin{tabular}[c]{@{}c@{}}Local host\\ (v3.1 Local)\end{tabular} & 0.4 \\ \cline{4-5} 
 &  &  & Physical & 0.2 \\ \cline{2-5} 
 & \multirow{3}{*}{\begin{tabular}[c]{@{}c@{}}Exploit \\ (v3.1 Attack)\\ Complexity\end{tabular}} & \multirow{3}{*}{\begin{tabular}[c]{@{}c@{}}EC \\ (AC)\end{tabular}} & Low & 1.0 \\ \cline{4-5} 
 &  &  & Moderate & 0.5 \\ \cline{4-5} 
 &  &  & High & 0.2 \\ \cline{2-5} 
 & \multirow{2}{*}{\begin{tabular}[c]{@{}c@{}}User \\ Interaction\end{tabular}} & \multirow{2}{*}{UI} & No & 1.0 \\ \cline{4-5} 
 &  &  & Yes & 0.3 \\ \cline{2-5} 
 & \multirow{3}{*}{\begin{tabular}[c]{@{}c@{}}\\ Authentication\\ (v3.1 - Privileges \\ Required)\end{tabular}} & \multirow{3}{*}{\begin{tabular}[c]{@{}c@{}}AU\\ (PR)\end{tabular}} & None & 1.0 \\ \cline{4-5} 
 &  &  & \begin{tabular}[c]{@{}c@{}}User\\  (v3.1 - Low)\end{tabular} & 0.6 \\ \cline{4-5} 
 &  &  & \begin{tabular}[c]{@{}c@{}}Admin \\ (v3.1 - High)\end{tabular} & 0.2 \\ \hline\hline
\end{tabular}
\end{table}

\subsubsection{Impact Metric}

In CVSS, the impact metric is used to evaluate different exploitation methods and capture the effects of successfully exploited vulnerabilities. This metric is determined using three factors:  \emph{confidentiality} ($C$), i.e., the effect on system information disclosure, \emph{integrity} ($In$), i.e., how detrimental the modification of system data would be, and \emph{availability} ($A$), i.e., the system accessibility after an adverse effect has occurred. During an attack, an adversary can cause \emph{high}, \emph{low}, or \emph{no impact} in each specified factor. For our study, the impact metric is designed to capture the effect of different exploited vulnerabilities in the EPS. During an attack on a $PV$ or a $PQ$ bus, the system may experience varying degrees of impacts related to \emph{total loss}, \emph{some loss}, or \emph{no loss} of the confidentiality, integrity, and availability of certain grid-supporting devices. More specifically, if the attacker is able to attack a $PV$ bus, we assume a worst-case scenario since the attacker demonstrated to have enough information and skills to attack a highly secure system and possibly have the means to exploit additional vulnerabilities. This, in turn, may result in a total loss of integrity, confidentiality, and availability. Using this assumption, the impact of compromising a $PV$ will cause \textit{high} impact on confidentiality, integrity, and availability. On the other hand, despite existing research demonstrating the importance of load altering attacks on power system stability \cite{amini2016dynamic}, manipulation of $PQ$ buses and load change attacks will likely not result in interruption of the operation of the generator, load, or transmission line in the system due to frequency load shedding protections \cite{huang2019not}. Under these circumstances, the impact of a compromised $PQ$ bus will not be significant enough when compared to the impact of a compromised $PV$ bus \cite{cao2020cyber}. Thus, we assume that compromising $PQ$ buses will have \textit{low} impact in all three categories. Finally, the \emph{no impact} value is used when an attacker compromises an electronic device that is not connected to any $PV$ or $PQ$ bus.

Based on the exploitability and impact metrics, CVSS can be calculated as shown in Eq. (\ref{eq:CVSS}) \cite{cvssreport}:

\begin{equation}
CVSS=
\left\{\begin{array}{l}
\text{min}~\{E+I, 10\}, \ \ ~~~~~~~\text{if} \  S \ \text{unchanged} \\
\text{min}~\{1.08 \cdot ({E}+{I}), 10\},~\text{if} \ S \ \text{changed}
\end{array}\right.
\label{eq:CVSS}
\end{equation}

\begin{equation} 
\label{eq:exploitability}
\begin{split}
{E}&= (AV \cdot AC \cdot UI \cdot PR ) \cdot 8.22 
\vspace{-0.15in}
\end{split}
\end{equation}

\begin{equation}
{I}=
\left\{\begin{array}{l}
6.42 \cdot I_b, \ ~~~~~~~~~~~~~~~~~~~~\text{if} \  S \ \text{unchanged} \\
7.52 \cdot (I_b -0.029) - \ \\  ~~~~~~ 3.25 \cdot (I_b-0.02)^{15}, \ \ \text{if} \ S \ \text{changed}
\end{array}\right.
\label{eq:impactmetric}
\end{equation}

\noindent where $I_{b}=1-[(1-C)\cdot(1-In)\cdot(1-A)]$. ${E}$, $AV$, $AC$, $UI$, $PR$, $S$, and ${I}$ represent the exploitability metric, attack vector, attack complexity, user interaction, privileges required, scope, and impact metrics, respectively. The calculated CVSS value is used as a major factor in the computation of the $TP$ within our transition model. 

The traditional CVSS scoring method provides a detailed calculation process that assesses the impact of exploiting a vulnerability with different attack vectors. However, it cannot be used directly for our application since it fails to consider important factors when used to evaluate complex cyberphysical systems. In particular for power systems, it does not take into account features such as system topology, power generation, and line constraints. Since we assume the adversarial agent does not have full topological information, we include power generation and line constraint calculations in our proposed $TP$ calculation. Since generators provide varying amounts of power to a system depending on the current state of the grid, the relative importance of a generator (and hence its attack impact) is determined by its power output. In addition to considering the difficulty of transitioning to certain system nodes, we also examine the overload percentage of the transmission lines. If the power flow across that line is near its flow constraint, the line could be more easily affected by changes in the surrounding system. Taking each of the aforementioned aspects into consideration, we define the $TP$ as follows in Eq. (\ref{eq:Trans}): 

\begin{equation} 
\label{eq:Trans}
\begin{split}
TP & = \frac{CVSS}{10} * \frac{G}{\sum_{k=1}^{n} G_{k}}* \frac{Pf}{\lambda^{critical}}
\end{split}
\end{equation}

\noindent where $G$ is the power generation of a connected generator, $n$ is the total number of generation units in the system, $Pf$ represents the power flow through transmission lines, and $\lambda^{critical}$ is the line power flow constraint for the connected transmission line. For a power system operating under normal conditions, the range of ${G}/{\{\sum_{k=1}^{n} G_{k}}\} \in [0,1]$ and $\lambda^{critical} \in [0,1]$. Since the $CVSS$ score $\in [0,10]$, we scale it by dividing by 10. A smaller $TP$ value represents a cyberphysical node vulnerability of low severity, i.e., the node has lower possibilities to be exploited by attackers since it has a lower CVSS score, and it is less important in terms of overload percentage, generation amount, and thermal limits. On the contrary, a $TP$ represents a cyberphysical node vulnerability of high severity. 

\subsection{ Step 4: Solution of Adversarial Model}
After formulating and defining the corresponding POMDP, in this step, we develop an algorithm to solve the model and yield the optimal transition policy for the considered threat. Due to the complexity of EPS, it is important to have a mechanism to solve sequential decision-making problems efficiently. In our studies, we develop a D$Q$N-based DRL algorithm. 

\subsubsection{$Q$-Learning}
$Q$-learning is an off-policy RL algorithm designed to find the optimal action an agent needs to take at the current state. All the actions that the RL agent can take are evaluated using a $Q$-value which determines how good is a particular action in the current state. Eq. (\ref{eq:q_learning}) shows the `update' rule used to calculate new $Q$-values at each state-action pair, where $s$ is current state, $a$ is the action at $s$, $s'$ is the new state, and $a'$ includes all the future potential actions. Using the learning rate $\alpha \in [0, 1]$, the new $Q$-value ($Q_{new}(s,a)$) is updated using the current $Q$-value ($Q_{old}(s,a)$), the estimated optimal future value ($\underset{a}{\max}Q(s',a')$), and the immediate reward ($R$). $\gamma \in [0,1]$ represents a discount factor that determines the importance of immediate rewards compared with potential long-term rewards. A higher $Q$-value demonstrates that a series of actions will produce a higher total accumulated reward. These actions are referred to as the \emph{optimal policy}.

\begin{multline}
 Q_{new}(s,a) = Q_{old}(s,a) \\
+~\alpha \cdot(R +~\gamma \cdot\underset{a}{\max}Q(s',a')-Q_{old}(s,a)) 
\label{eq:q_learning}
\end{multline}

Traditionally, $Q$-learning is implemented using $Q$-tables. However, this approach is not practical nor scalable for solving large state-action environments. To solve this issue, researchers in \cite{mnih2013playing} proposed the replacement of $Q$-tables with deep neural networks, also known as D$Q$Ns.

\subsubsection{Deep $Q$-Network (D$Q$N)}

\begin{figure}[t]
\centerline{\includegraphics[width=0.3\textwidth]{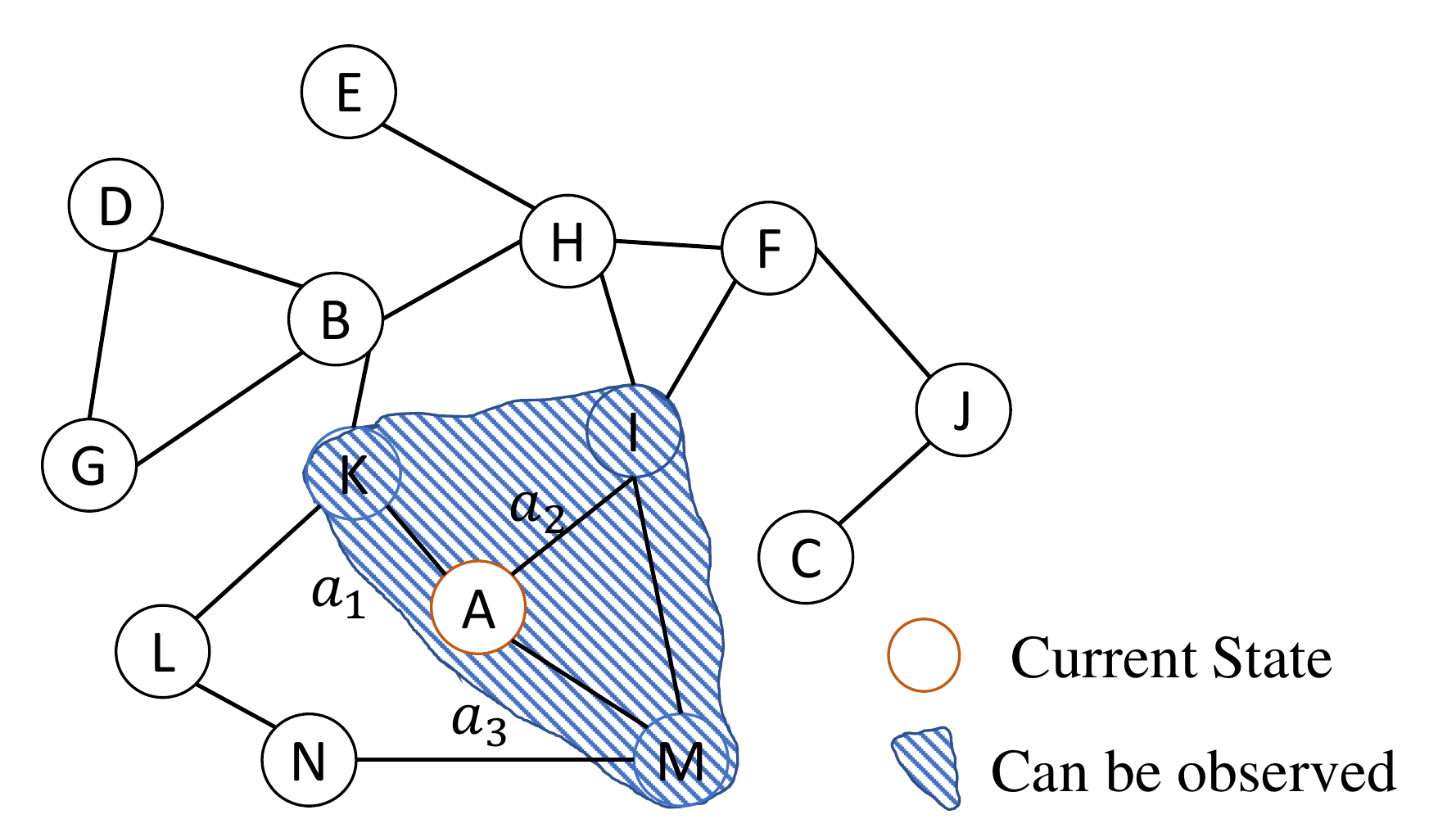}}
\caption{Example of the observation process for state A.} 
\label{fig:Observation}
\end{figure}

In order to address the computational overhead of $Q$-learning when dealing with large, uncertain, and dynamic environments, D$Q$Ns generalize the approximation of the $Q$-value function using artificial neural networks rather than storing every solution in a table. For our environment modeled as a POMDP, we assume that the D$Q$N agent starts in a random initial state $s$ (a node in the cyberphysical network) and transitions to the next state $s'$ occur by taking actions (i.e., moving through nodes/buses in the system) based on observations until it reaches both nodes of the contingency pair, regardless of transition order. As shown in Eq. (\ref{eq:o}), an observation function $O$ generates the observations for each potential action $a^{\prime}$ given state $s$. The state $s$ refers to the bus where the agent is currently located during the solution process {including all bus-related corresponding information and measurements (i.e., circuit breaker status, power generation, power consumption, voltages, etc.)}. At every step, the agent makes an observation $o$ which helps to indicate which possible movements (or actions) the agent can take according to the observed available neighboring buses and branches connected to the current bus. Then, based on the observation, an action $a$ (movement to another bus) is performed in order to transition to the next state (i.e., the next bus). Fig. \ref{fig:Observation} presents an overview of this process. It can be seen that if the current state of the agent is bus A, then the attacker (agent) can make an observation $o$ to obtain the next available bus to move into, which can be either K, I, or M, by taking $a_{1}$, $a_{2}$, or $a_{3}$, respectively. Using the observation, the agent is also capable of revealing whether or not a contingency is present in the current bus where it moved into, since the contingency pairs are known to the attacker from the analysis in step 2 (Section \ref{ss:contingencystep2}).

\begin{equation}
\label{eq:o}
O(o |s, a)=O(o | s^{\prime}, a^{\prime}) 
\end{equation}

Once each potential transition is determined, the $TP$ for each transition needs to be computed (as defined in Eq. (\ref{eq:Trans})). These calculated results will be utilized to determine the security index, $SI'_i$, when making a transition from $s$ to $s'$ as shown in Eq. (\ref{eq:security}), where $\gamma$ is a discount factor, and $\Delta C_{p}$ is the line flow difference between the current state and each potential transition state. Finally, the maximum value of the security index which represents the bus with the highest vulnerabilities' score and overload value, will be used to compute the corresponding reward function. As shown in Eq. (\ref{eq:reward}), the reward function considers the overall benefit of different transitions as it takes into account the security index of each potential state, $SI'_i$.

\begin{equation}
SI'_{i} = \underset{a\in \mathcal{A}}{\max}\ {\gamma\cdot\underset{s'\in \mathcal{S}}{\sum}TP\left(s'|s, a\right) \cdot\Delta C_{p}}
\label{eq:security}
\end{equation}

\begin{equation}
R=\underset{s'\in \mathcal{S}}{\sum} TP\left(s'|s, a\right) \cdot[\Delta C_{p}+SI'_{i}])
\label{eq:reward}
\end{equation}

State-action-rewards tuples are stored in the replay memory set $M$ for recording agent's experiences. This memory set assists in independently training the neural network. All environmental information of the current state (weights, biases) is stored in the action-value parameter $\theta$. In each step, the D$Q$N combines multi-layered neural networks with existing $Q$-learning algorithms to approximate $Q(s,a;\theta)$. $\theta^{-}$ will change as the result of changing $\theta$. Eq. (\ref{eq:target}) demonstrates the updated target value given by the current state and action, where the target action-value parameter $\theta^{-}$ is equal to $\theta$ at the beginning of the iterations. When this number of iterations is reached during training, $\theta^{-}$ is updated to prevent an obstructed learning process \cite{mnih2015human}. Using the parameters described, the loss-function value can be calculated as shown in Eq. (\ref{eq:loss}) for each state-action pair. It represents the error between the predicted $Q$-value and the target $Q$-value. The goal is to determine an \emph{optimal policy} that minimizes the error and ensures that the training result will be as close as possible to the target value, where the target value is the estimated expected return of the actions taken by the D$Q$N.

\begin{equation}
\label{eq:target}
y_{i}=R+\gamma \max _{a^{\prime}} {Q}\left(s^{\prime}, a^{\prime} ; \theta^{-}_{i}\right)
\end{equation}

\begin{equation}
\label{eq:loss}
\mathcal{L}_{i}\left(\theta_{i}\right)=\mathbb{E}_{\left(s, a, R, s^{\prime}\right) \sim \mathcal{M}}\left[\left(y_{i}-Q\left(s, a ; \theta_{i}\right)\right)^{2}\right]
\end{equation}

\begin{figure}[t]
\centerline{\includegraphics[width=0.5\textwidth]{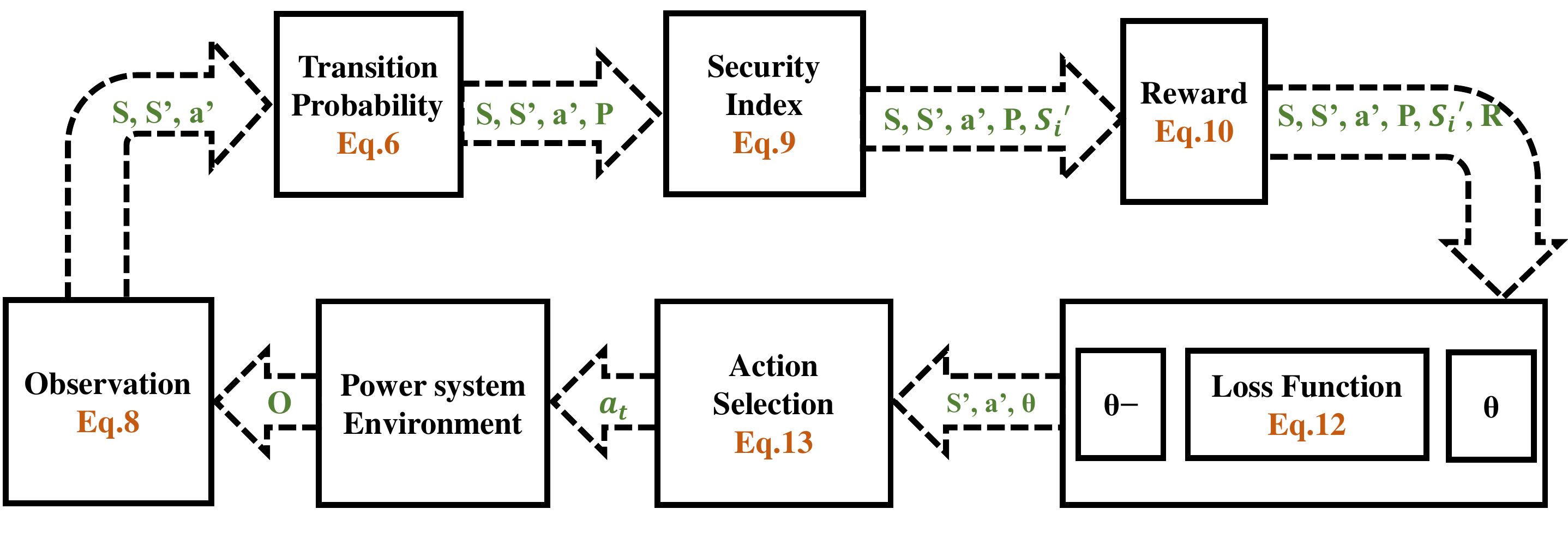}}
\caption{Process of how the D$Q$N model is utilized as part of the proposed cybersecurity assessment.} 
\label{fig:Process}
\end{figure}

The agent performs an action that is selected according to the designated exploration-exploitation ($\epsilon$-greedy) strategy of Eq. (\ref{eq:epsilon}). Such a strategy controls the degree of exploitation over exploration. At each step, if exploration is being performed with probability $\epsilon$, the algorithm selects a random action $a_t$ from the action set. During exploitation with probability $1-{\epsilon}$, the action with the maximum $Q$-value is taken. The target values $\theta^{-}$ will only be updated once the desired number of iterations has been reached \cite{mnih2015human}. The overall learning process is presented in Fig. \ref{fig:Process} and the notation of the D$Q$N parameters is summarized in Table \ref{table:dqnparameters}. This learning process is repeated until a terminal state is reached, i.e., both contingency pairs have been finally ``visited'' by the agent.

\begin{equation}
a_{t} = \begin{cases}  random \ a^{\prime}, &  \epsilon \\ 
\arg max_{a} Q(s^{\prime},a^{\prime};\theta), & 1- \epsilon \\
\end{cases}
\label{eq:epsilon}
\end{equation}

\begin{table}[]
\centering
\caption{Notation of D$Q$N parameters.}
\label{table:dqnparameters}
{
\begin{tabular}{||c|c||}
\hline\hline
\multicolumn{2}{||c||}{\textbf{DQN parameters}} \\ \hline\hline
$\alpha$ & Learning rate  \\ \hline
$\gamma$ & Discount factor\\ \hline
$s$ &  Current node\\ \hline
$s'$ & Next node transitioned from $s$ \\ \hline
$TP $& Transition probability \\ \hline
$\Delta {C}_{p} $& Line flow difference between $s$ and $s'$ \\ \hline
$SI'$ & Security index \\ \hline
$R$ & Reward \\ \hline
$\theta$ & Action-value parameter \\ \hline
$\theta^{-}$ & Target action-value parameter  \\ \hline
\hline
\end{tabular}}
\end{table}

\subsection{Step 5: Output of the Assessment Process}

An attacker with sufficient OSINT can aggregate enough power system information (e.g., power generation, capacity, load consumption, topological data, etc.) to perform contingency analysis and identify critical system elements. These identified critical contingency elements can be leveraged to generate cyberattack transition policies following the process described in previous steps. The generated cyberattack transition policies take into account vulnerabilities in electronic devices that exist in the cyber network layer as well as physical system vulnerabilities related to contingency studies. The DRL algorithm provides a solution known as the optimal attack transition policy that can be used to attack the devices controlling the operations of the critical elements (e.g., microprocessor-based relays controlling circuit breakers, protocol translator converters, etc.) and results in potential power outages in the EPS. Our methodology can also be leveraged by control center operators and stakeholders to identify vulnerable components in the EPS or investigate potential attack strategies.

\section{Contingency Analysis Simulations}\label{s:result}
In this section, we introduce a number of contingency simulation case studies used to demonstrate the effectiveness of the proposed approach. These case scenarios prove how the most critical contingency pairs of a system vary when wind energy systems are in-place. We provide an analysis of the varying degrees of severity with different contingency scenarios and examine how wind generation impacts critical contingencies. For this validation study, we use a doubly-fed induction generator (DFIG) model for wind power generation modeling and real-time simulation (OPAL-RT) for testing the system in a real-time environment.

\subsection{Contingency Scenarios}

First, we run the assessment process of Section \ref{s:method} up to \textit{Step 2} in order to assess multiple contingency scenarios in different test systems. In Table \ref{tab:Numberofcontingency}, we present the number of critical contingencies for $N-1$, $N-1-1$, and $N-2$ scenarios in different power system test cases. For example, the IEEE 39 bus system has 13 $N-1$, 19 $N-1-1$, and 71 critical $N-2$ contingencies without any wind power injection, while the number of these contingencies varies with different wind penetration levels. 
\begin{table}[t]
\caption{Number of contingencies for different cases and scenarios.}
\begin{center}
\begin{tabular}{||c|c|c|c||}
\hline
\hline
{Case Name} &{$N-1$}&{$N-1-1$}&{$N-2$}  \\
\hline\hline
{IEEE 30 Bus System} &{1}&{2}&{8}\\
\hline
{IEEE 39 Bus System} &{13}&{19}&{71} \\
\hline
IEEE 39 Bus System + &\multirow{2}{*}{16}&\multirow{2}{*}{24}&\multirow{2}{*}{73}\\

Wind (Table \ref{tab:WindContingency1}: SC7, $t=800m$)&&&\\
\hline

IEEE 39 Bus System + &\multirow{2}{*}{17}&\multirow{2}{*}{26}&\multirow{2}{*}{67}\\
Wind (Table \ref{tab:WindContingency1}: SC7, $t=1400m$)&&&\\
\hline

IEEE 39 Bus System + &\multirow{2}{*}{16}&\multirow{2}{*}{23}&\multirow{2}{*}{103}\\
Wind (Table \ref{tab:WindContingency2}: SC5, $t=0m$)&&&\\
\hline

{UIUC 150 Bus System} &{176}&{174}&{442} \\
\hline
{Polish 2383 Bus System} &{2236}&{2234}&{15881} \\

\hline\hline
\end{tabular}
\label{tab:Numberofcontingency}
\end{center}
\end{table}

\begin{figure}[t]
\centerline{\includegraphics[width=0.9\linewidth]{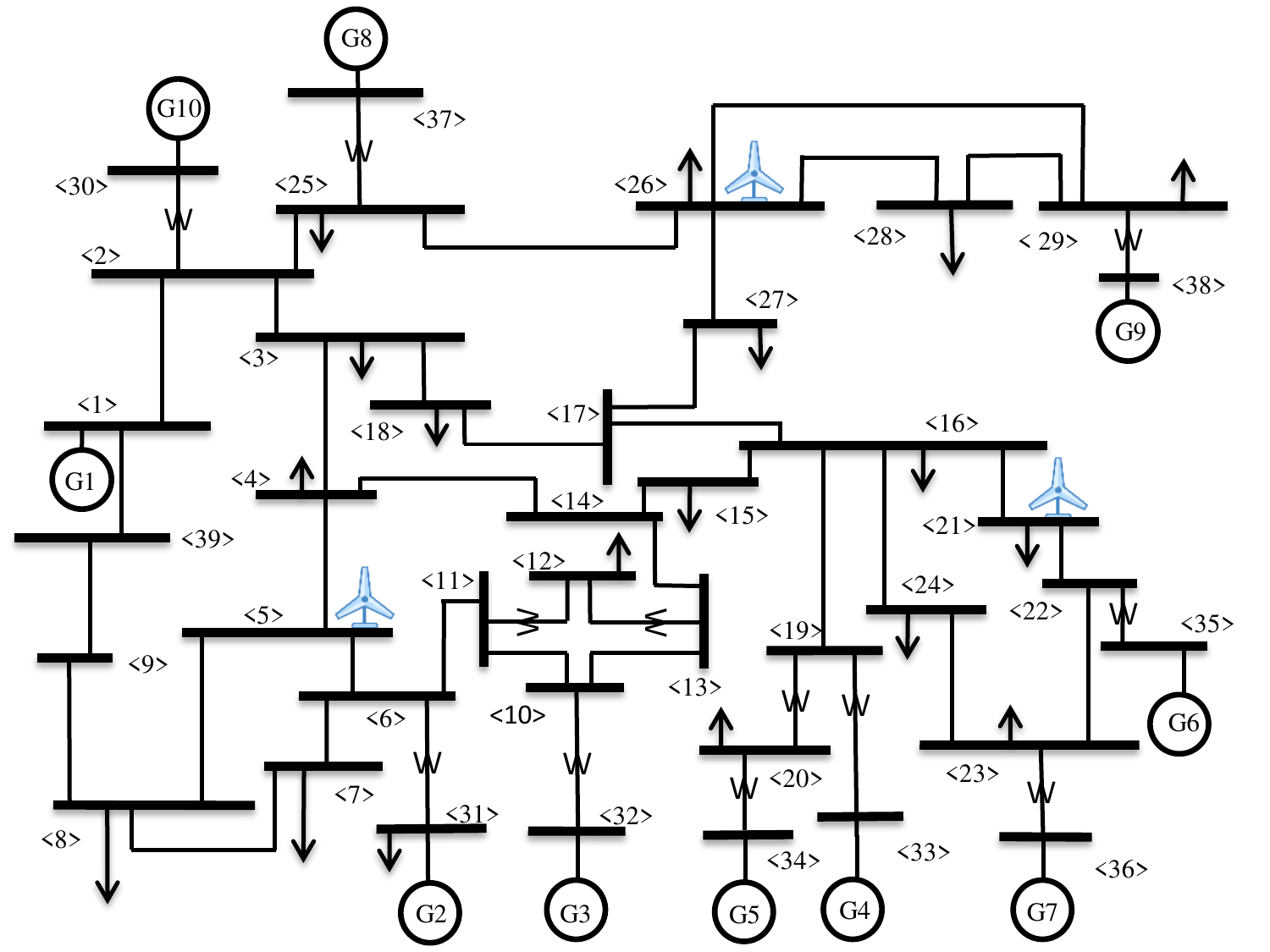}}
\caption{IEEE 39 bus system with wind power integration.} 
\label{fig:39}
\end{figure}

The $N-1$ contingencies are determined by disconnecting each line and observing system responses. For $N-1-1$, the most severe $N-1$ case is removed from the system, and the process is run again. The $N-2$ pruning algorithm is carried out as described in Section \ref{s:method}. In the rest of the section, we focus on the $N-2$ case as the most severe scenario. It should be noted that the proposed approach can be adapted, based on user requirements, for any number of contingencies $k$.

\subsection{Wind Power Generation Modeling using a DFIG Model}\label{ss:DFIG}

\begin{figure}[t]\centering 
\subfigure[] { \label{fig:winddata1}
\includegraphics[width=0.8\linewidth]{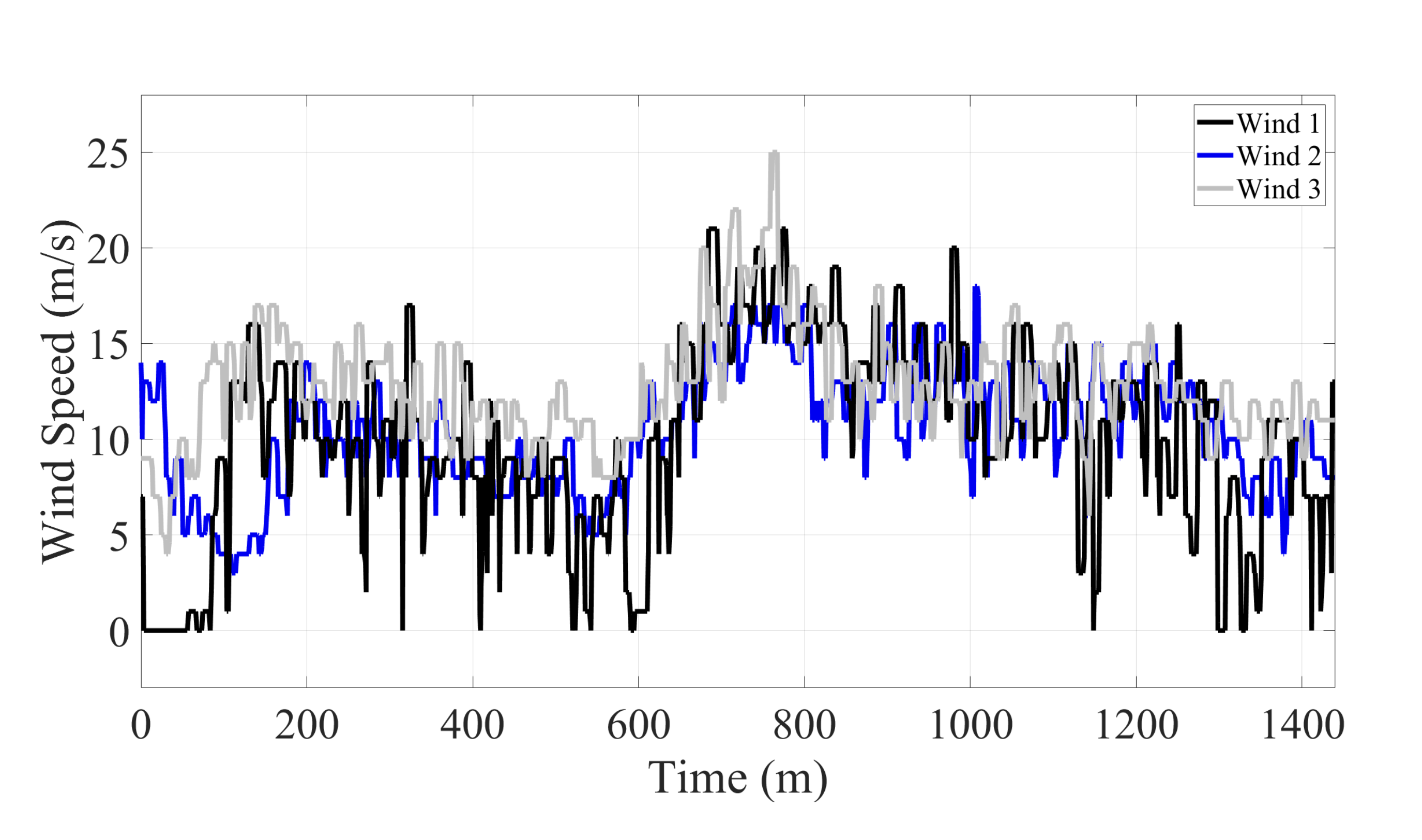}}
\vspace{-3mm}
\subfigure[] { \label{fig:winddata2} 
\includegraphics[width=0.8\linewidth]{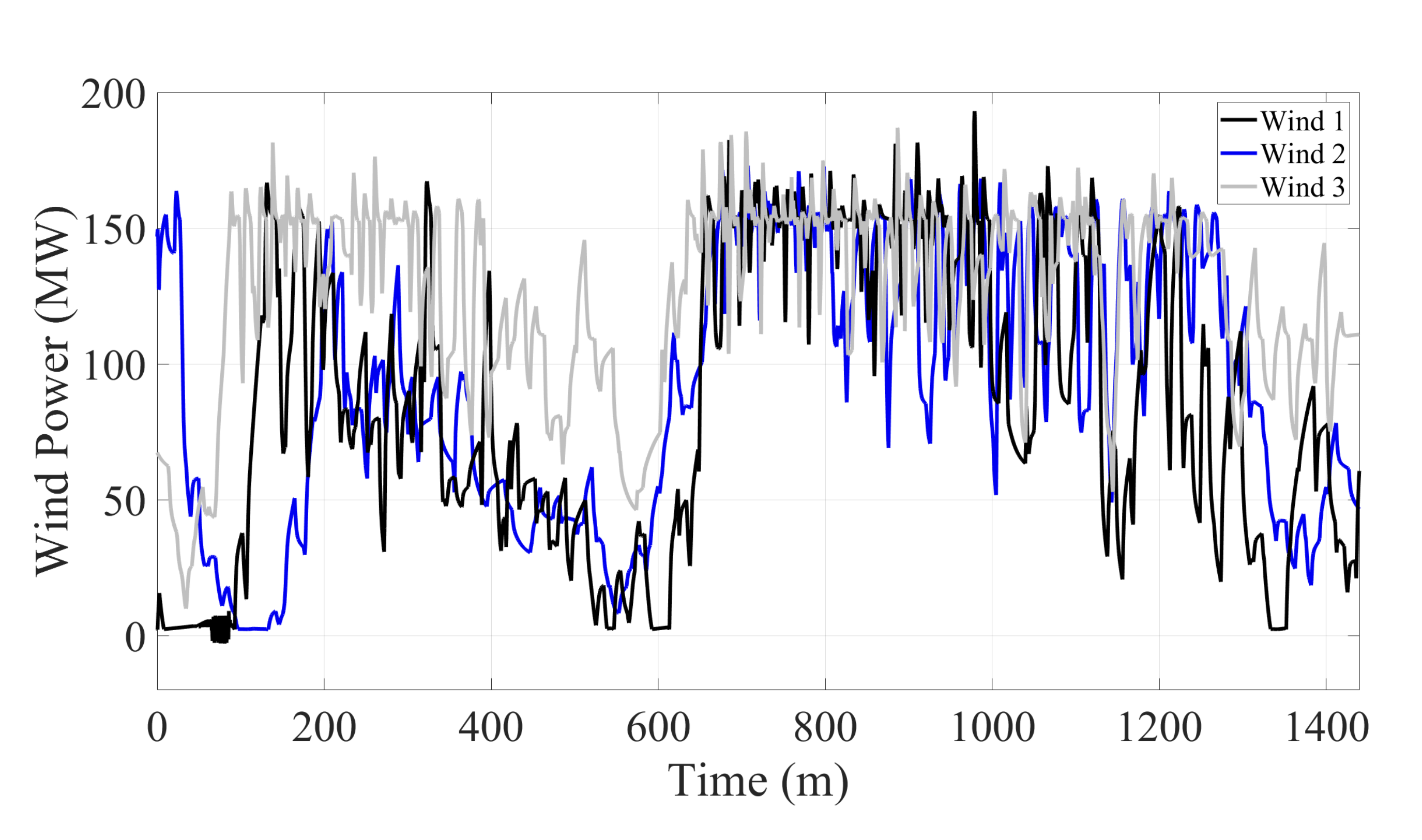}}
\caption{\subref{fig:winddata1} Wind speed, and \subref{fig:winddata2} wind power variation for \emph{scenario A}.}
\end{figure}
\begin{figure}[t]\centering 
\subfigure[] { \label{fig:winddata3}
\includegraphics[width=0.8\linewidth]{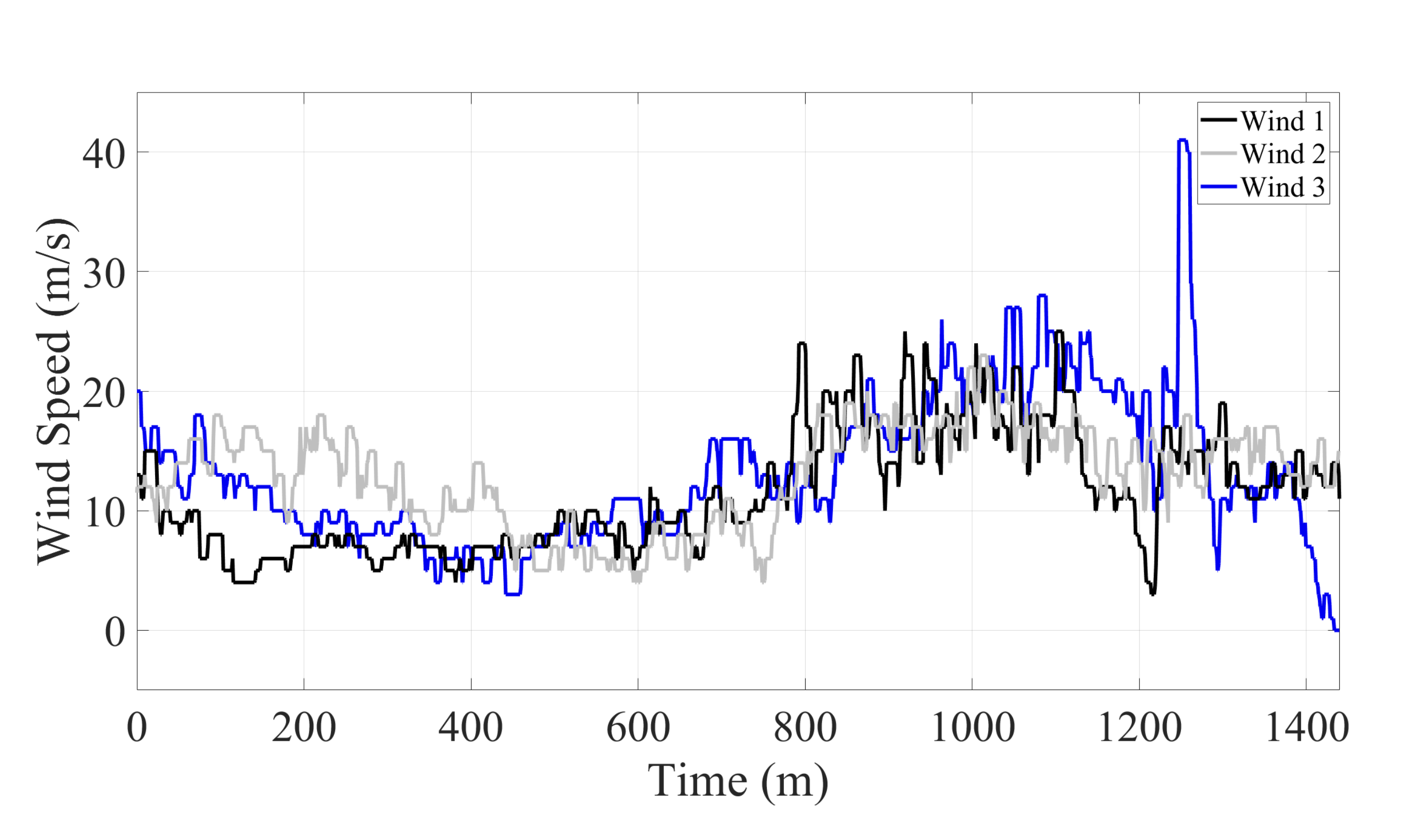}}
\vspace{-3mm}
\subfigure[] { \label{fig:winddata4} 
\includegraphics[width=0.8\linewidth]{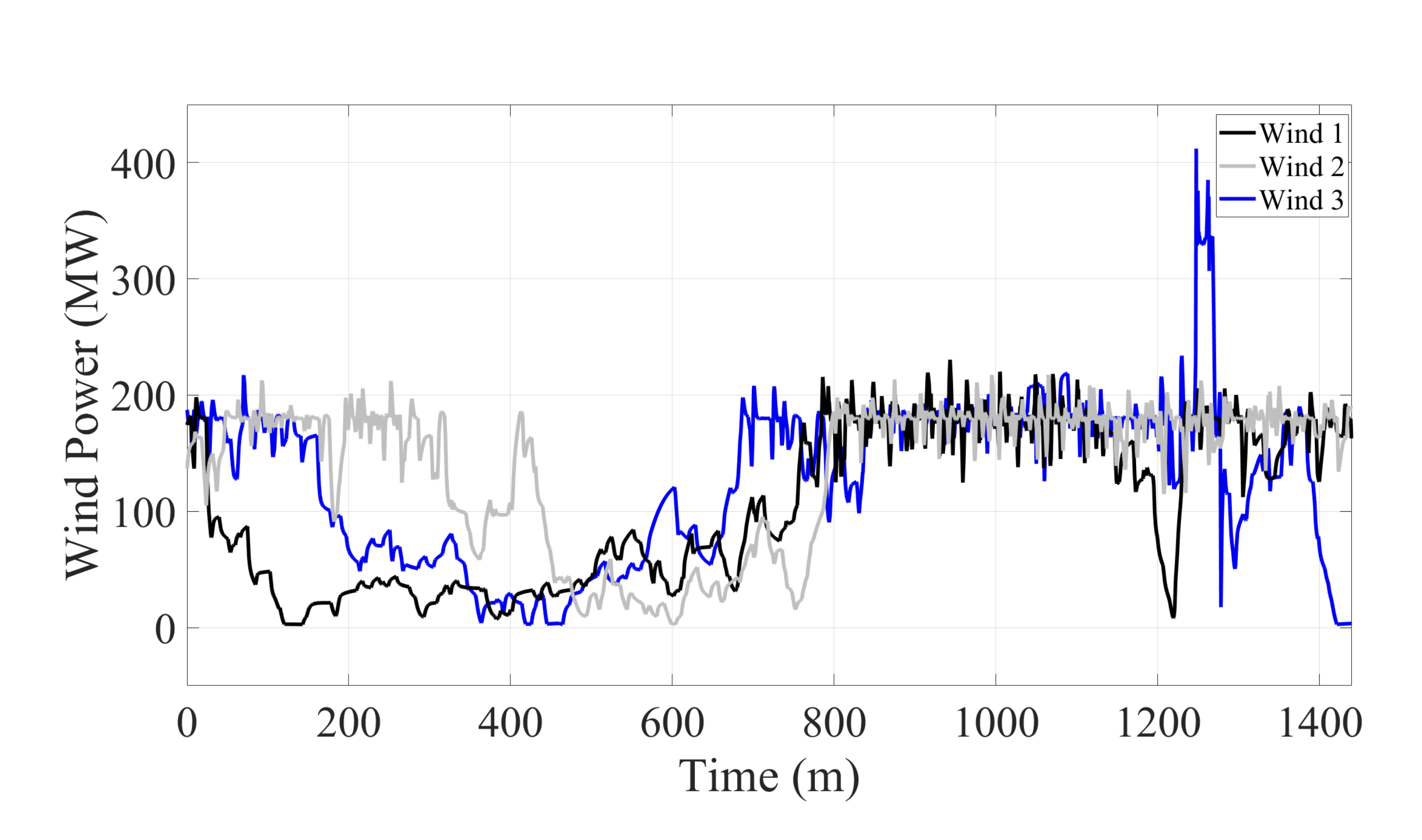} }
\caption{\subref{fig:winddata3} Wind speed, and \subref{fig:winddata4} wind power variation for \emph{scenario B}.}
\end{figure}

\begin{table*}[ht]
\caption{Scenario A: Impact of different wind power injections on the number of $N-2$ contingencies in the IEEE 39 bus system.}
\begin{center}
\begin{tabular}{||c|c|c|c||c|c|c|c|c|c|c||}
\hline\hline
Time&WF1&WF2&WF3&SC1&SC2&SC3&SC4&SC5&SC6&SC7 \\

(m)&(MW)&(MW)&(MW)&WF1&WF2&WF3&WF$1$+WF$2$&WF$2$+WF$3$&WF$1$+WF$3$&WF$1$+WF$2$+WF$3$ \\\hline\hline

{0} &2.39 &146.90 &67.30&71 &78 &79& 78& 85&80&84\\
\hline

{200} &108.10 & 131.10 & 125.60&52 &77&83&64 &92&65&80\\
\hline

{400} &90.92&54.10 &96.80&51 &74 &81&52 &83 &62&72\\
\hline

{600} &2.68 &54.79 &73.68 &71 &74 &79&76&82&80&83\\
\hline

{800} &153.70 &159.20 &155.70 &53 &77 &84&71 &\cellcolor{gray}100 &78&73\\
\hline

{1000} &93.11 &76.37 &148.40 &52 &76 &83&51 &91 &67&82\\
\hline

{1200} &154.30&117.50 &153.80&53 &77&83&68 &96&78&79\\
\hline

{1400} &77.52 &54.52 &105.70&81 &74 &81&\cellcolor{gray}50 &83 &63&67\\
\hline

\hline
\hline
\end{tabular}
\label{tab:WindContingency1}
\end{center}
\end{table*}

\begin{table*}[ht]
\caption{Scenario B: Impact of different wind power injections on the number of $N-2$ contingencies in the IEEE 39 bus system.}
\begin{center}
\begin{tabular}{||c|c|c|c||c|c|c|c|c|c|c||}
\hline\hline
Time&WF1&WF2&WF3&SC1&SC2&SC3&SC4&SC5&SC6&SC7 \\

(m)&(MW)&(MW)&(MW)&WF1&WF2&WF3&WF$1$+WF$2$&WF$2$+WF$3$&WF$1$+WF$3$&WF$1$+WF$2$+WF$3$ \\\hline\hline

{0} &174.30 &136.80 &187.30&55 &77 &91& 73& \cellcolor{gray}103&85&68\\
\hline

{200} &30.66 & 178.00 & 66.20&74 &80&80&81 &85&82&87\\
\hline

{400} &15.39&107.00 &29.09&69 &77 &72&78 &79 &77&80\\
\hline

{600} &27.94 &3.56 &119.20 &74 &71 &82&74&82 &86&86\\
\hline

{800} &179.20 &182.90 &134.80 &56 &80 &83&80 &97 &79&67\\
\hline

{1000} &176.80&178.20 &180.20 &55 &80 &89&81 &101 &86&71\\
\hline

{1200} &75.87&161.20 &178.40&81 &78&89&76 &100 &70&90\\
\hline

{1400} &126.30 &163.90 &69.06&\cellcolor{gray}52 &79 &79&70 &87 &58&81\\
\hline

\hline
\hline
\end{tabular}
\label{tab:WindContingency2}
\vspace{-2mm}
\end{center}
\end{table*}

A DFIG model consists of a wound rotor induction generator driven by wind turbines and an AC/DC/AC insulated-gate bipolar transistor-based pulse width modulated converter. The DFIG model used in our case studies for modeling wind energy systems is developed in MATLAB/Simulink. Using this model, we are able to study the dynamic response of EPS to wind speed variations and investigate the impact of different penetrations. Three DFIGs are modeled and integrated to the IEEE 39 bus system at buses 5, 21, and 26 (Fig. \ref{fig:39})  \cite{bevrani2010intelligent}. The wind speed and wind power data for each wind system are collected at a one-minute resolution on May 14, 2020 (1440 mins = 24 hrs) from \cite{winddata}. In the rest of the paper, we investigate two scenarios of wind integrated power systems: \emph{scenario A}, in which the wind data is collected from three locations in Tallahassee, FL, with similar variation and power generation levels. The wind speed and corresponding wind power generation information are provided in Fig.\ref{fig:winddata1} and Fig.\ref{fig:winddata2}. In \emph{scenario B}, the wind speed and power data are obtained from three locations in Boston, MA (Wind 1), Dallas, TX (Wind 2), and Tiffin, OH (Wind 3) with different weather characteristics. For this scenario, the wind speed and power generation are shown in Fig.\ref{fig:winddata3} and Fig.\ref{fig:winddata4}, respectively.

\subsection{Contingency Scenarios with Wind Power Injection}
\begin{figure*}[t]\centering
\subfigure[] { \label{fig:a1}     
\includegraphics[width=8cm]{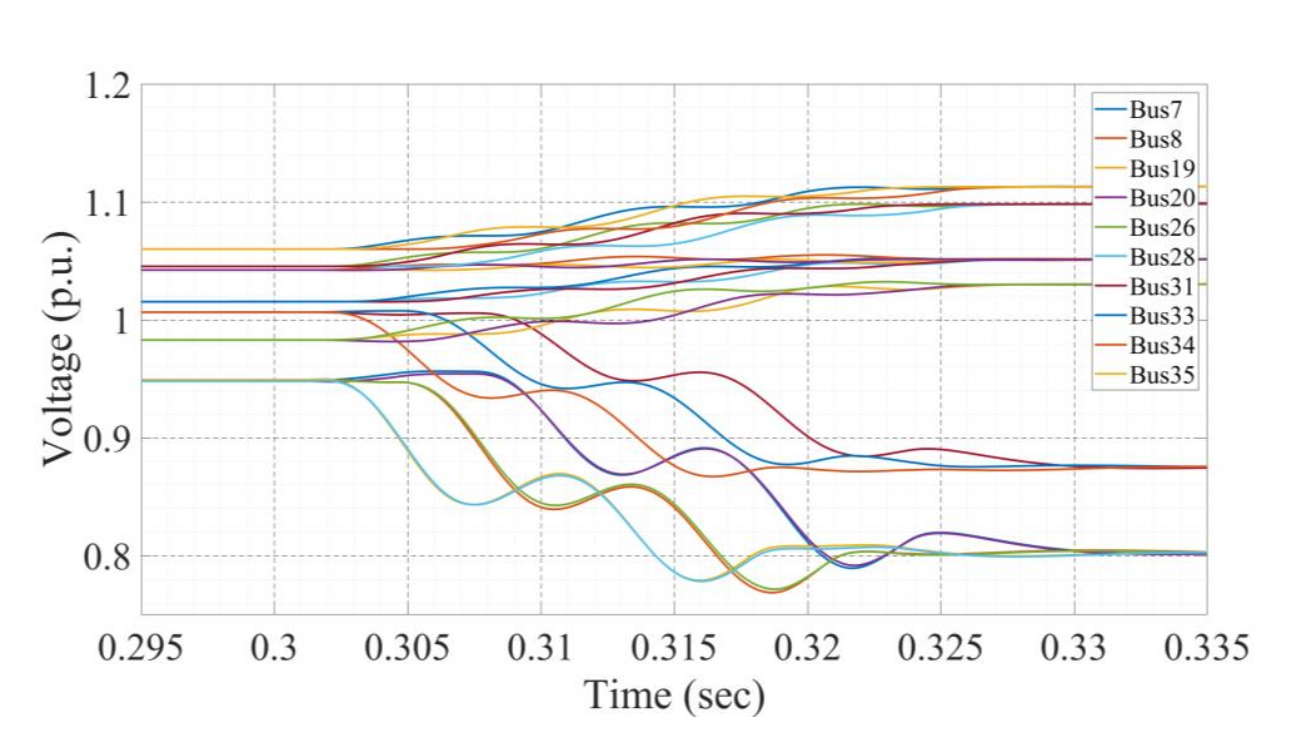}}
\subfigure[] { \label{fig:a2}     
\includegraphics[width=8cm]{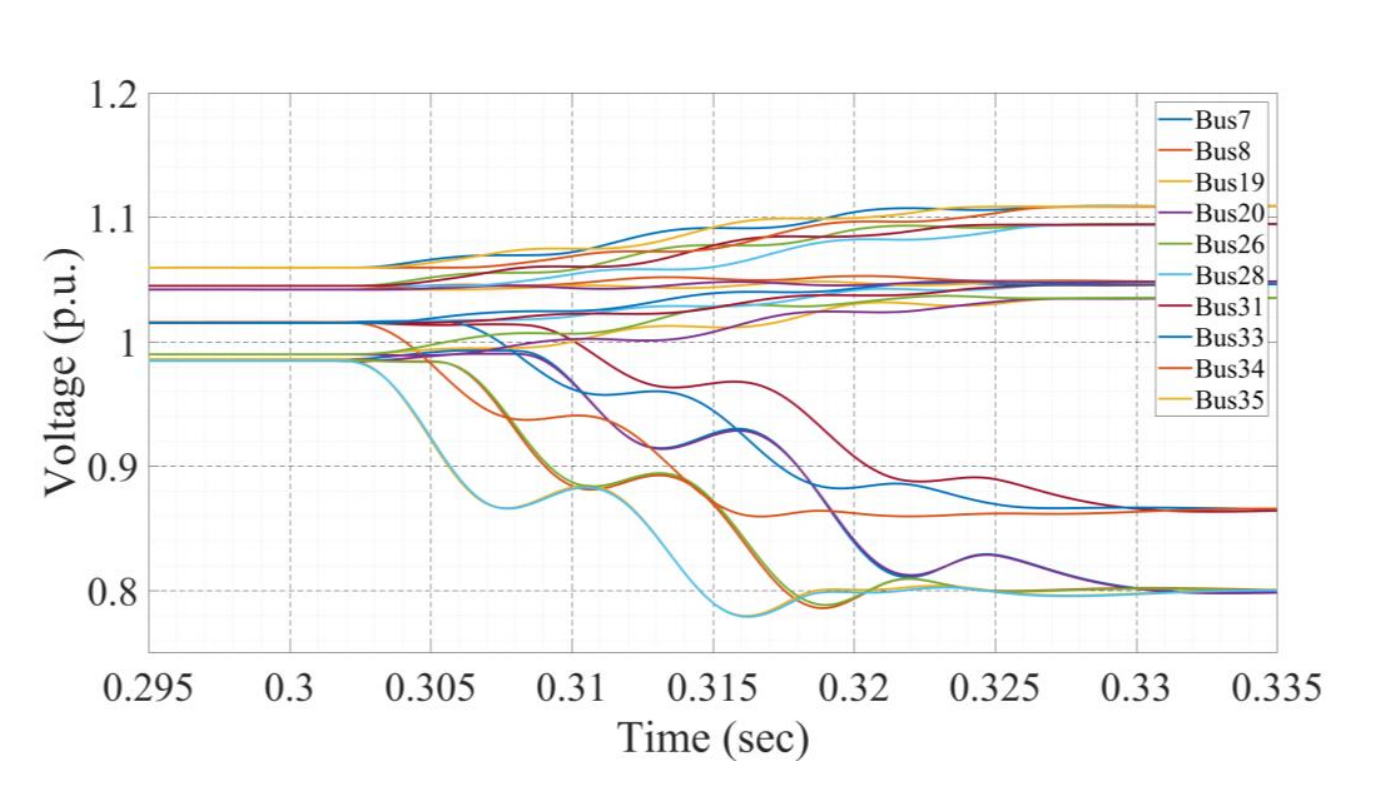}}\\
\subfigure[] { \label{fig:a3}     
\includegraphics[width=8cm]{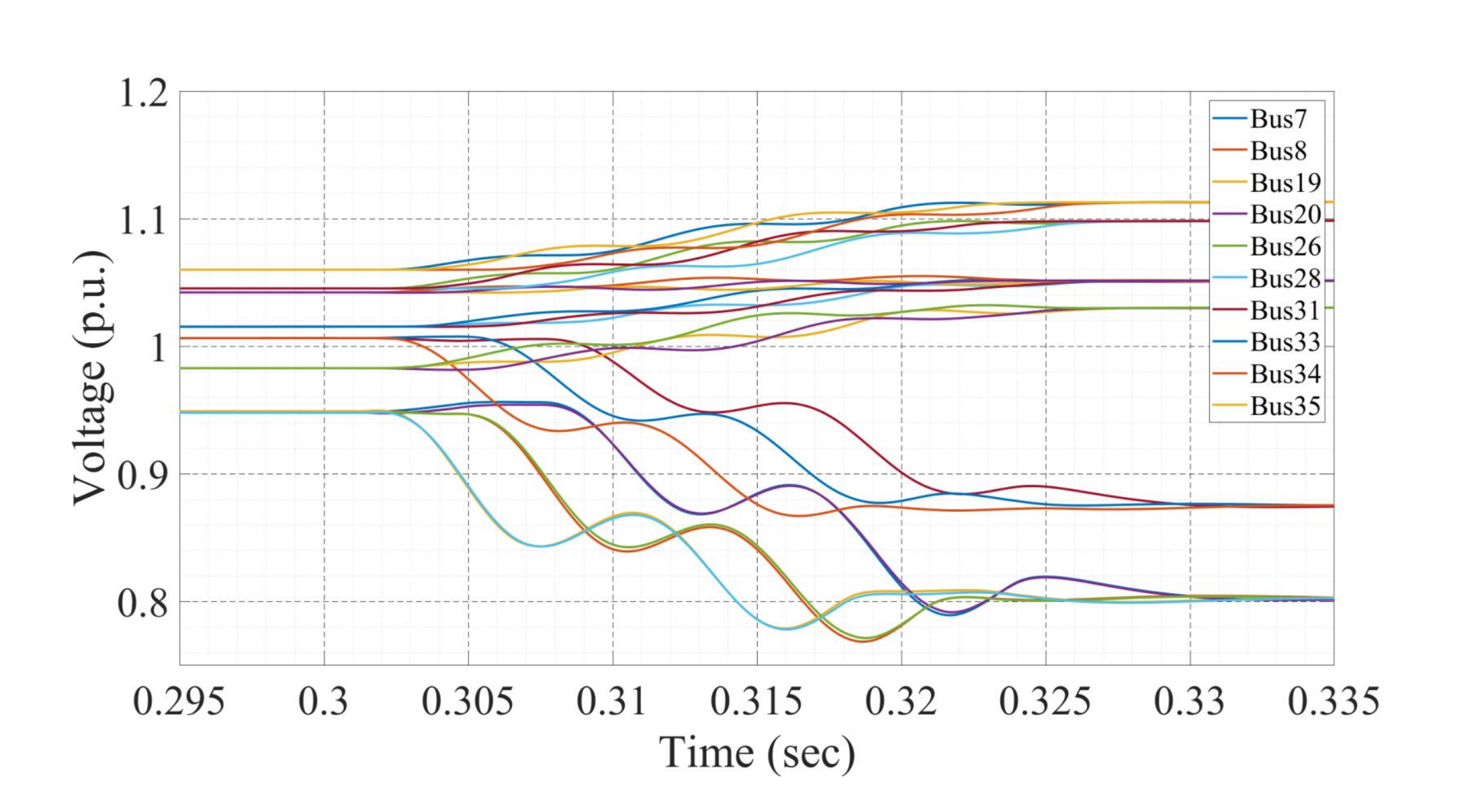}}
\subfigure[] { \label{fig:a4}     
\includegraphics[width=8cm]{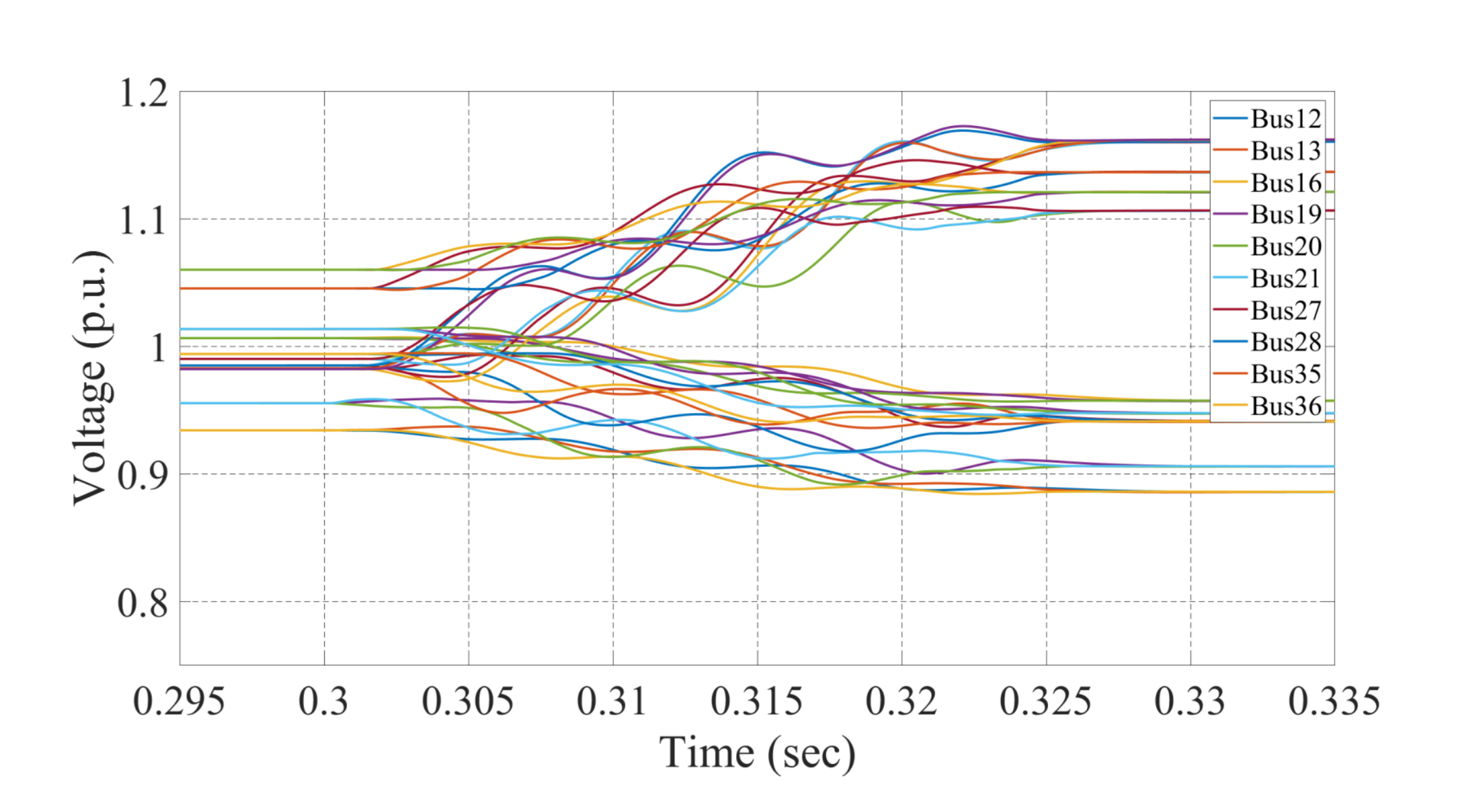}}\\
\subfigure[] { \label{fig:a5}     
\includegraphics[width=8cm]{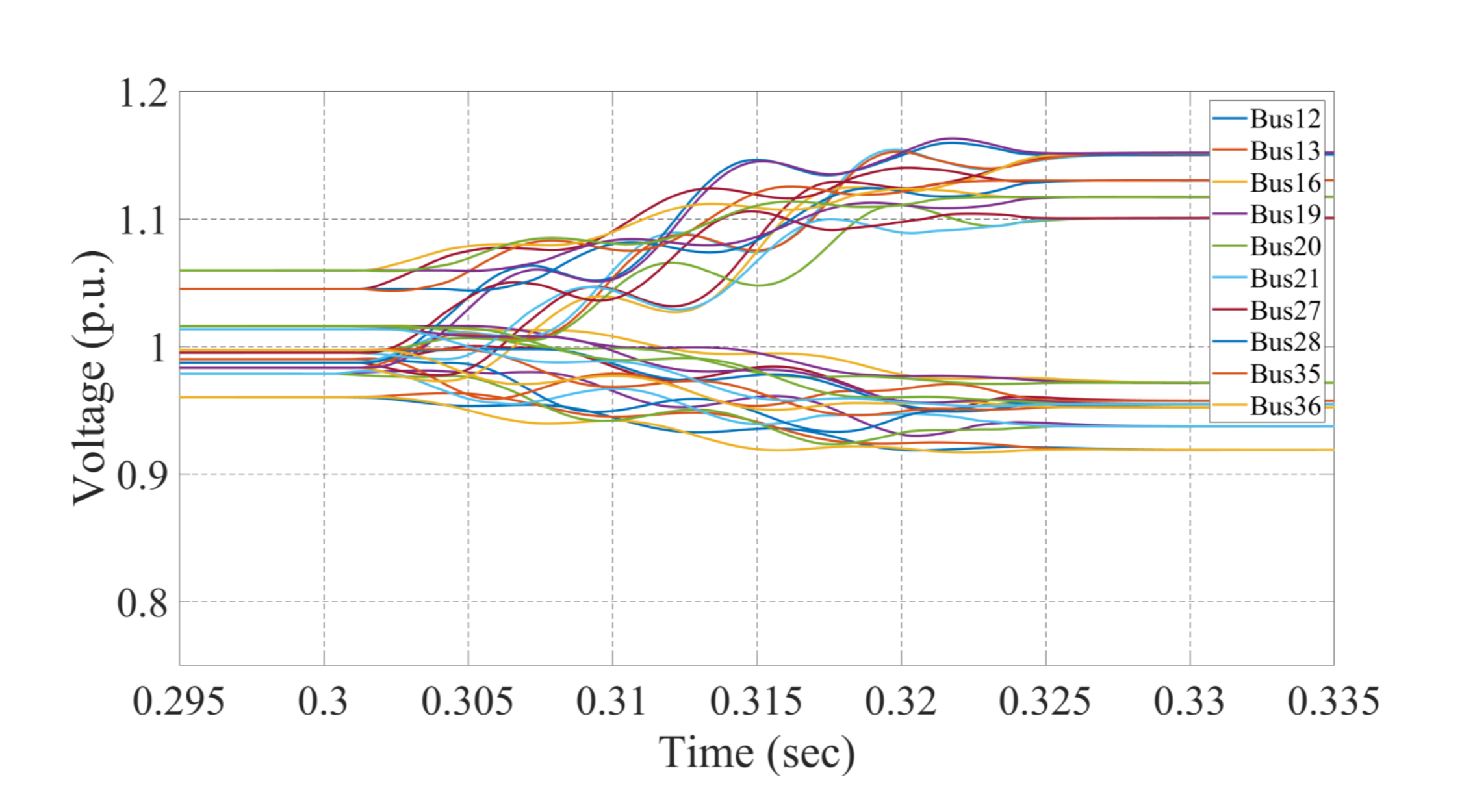}}
\subfigure[] { \label{fig:a6}     
\includegraphics[width=8cm]{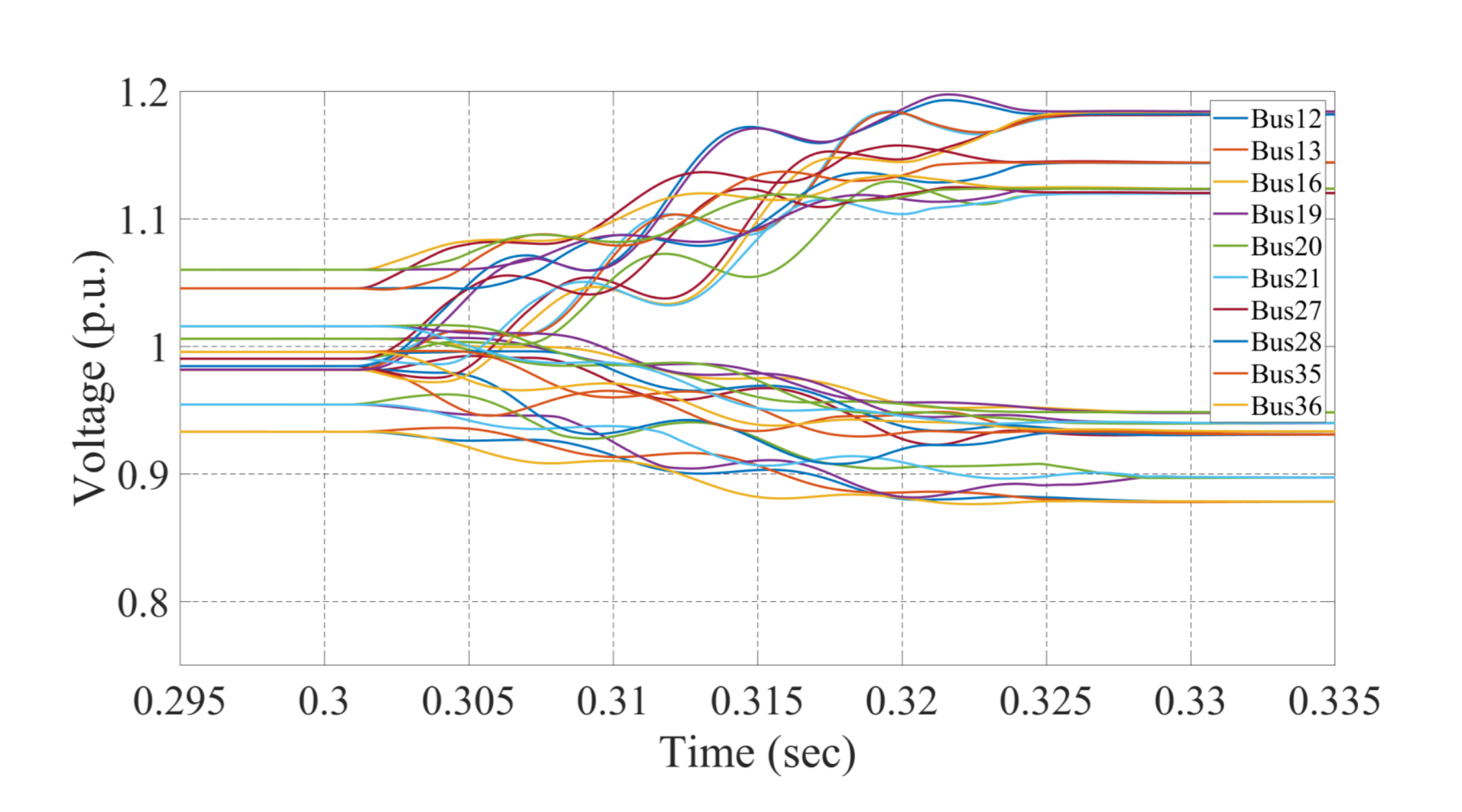}}

\caption{Contingency scenarios for IEEE 39 bus system: \subref{fig:a1} contingency pair at lines $5-8, 6-7$ without wind penetration, \subref{fig:a2} contingency pair $5-8, 6-7$ with wind penetration levels as case of Table \ref{tab:WindContingency1}: SC7, $t=800m$, \subref{fig:a3} contingency pair $5-8, 6-7$ with wind penetration as case of Table \ref{tab:WindContingency2}: SC5, $t=0m$. \subref{fig:a4} contingency pair $10-13, 16-21$ without wind penetration, \subref{fig:a5} contingency pair $10-13, 16-21$ with wind penetration as case of Table \ref{tab:WindContingency1}: SC7, $t=800m$, \subref{fig:a6} contingency pair $10-13, 16-21$ with wind penetration as case of  Table \ref{tab:WindContingency2}: SC5, $t=0m$.}
\label{fig:results}
\end{figure*}

The amount of power produced by wind energy systems fluctuates due to wind's intermittent nature. As the generation changes, power flow varies, which may affect contingency analysis results. Therefore, we simulate wind power injection levels at eight distinct timestamps, for the two simulation scenarios (\emph{scenario A} and \emph{scenario B}) throughout one day and observe the changes in reported contingencies with different wind penetration. These tests are performed for the IEEE 39 bus system (Fig. \ref{fig:39}). As shown in Table \ref{tab:WindContingency1} and Table \ref{tab:WindContingency2}, seven wind power integration simulation cases (SC1-SC7) are simulated for \emph{scenario A} and \emph{scenario B}. For each case, we present the amount of power injected by the three DFIG-based wind farms (WF$1$, WF$2$, WF$3$) and the number of identified $N-2$ contingencies. 

For \emph{scenario A} in Table \ref{tab:WindContingency1}, the highest number of $N-2$ contingency pairs ($100$) exists when WF2 and WF3 are integrated to the system (SC$5$) with generation of $159.20$MW and $155.70$MW, respectively. The least amount of pairs occurs when WF1 and WF2 turbines are injecting power into the system (SC$4$), and the wind power injection for WF1 and WF2  are $77.52$MW and $54.52$MW, respectively. As shown in the results, the number of $N-2$ contingencies change when the same amount of power is injected at different locations. Additionally, injecting varying levels of power in the same location also changes the number of contingencies. 

As for \emph{scenario B} in Table \ref{tab:WindContingency2}, the highest number of $N-2$ contingency pairs ($103$) exists when WF2 and WF3 are integrated into the system (SC$5$), and the wind power injection for WF2 and WF3 are $136.80$MW and $187.30$MW, respectively. The least amount of pairs occurs when only WF1 is injecting $126.30$MW power into the system (SC$1$). 

Comparing with the normal case of IEEE 39 bus system without wind power injections ($71$ pairs of $N-2$ contingencies in  Table \ref{tab:Numberofcontingency}), the number of $N-2$ contingencies in $38$ cases (out of $56$ cases in total in Table \ref{tab:WindContingency1}) of \emph{scenario A} are over $71$. For \emph{scenario B}, $44$ cases (out of $56$ cases in total in Table \ref{tab:WindContingency2}) are more than $71$. These results demonstrate how the intermittent behavior of wind energy directly affects the number and location of contingencies in EPS with high penetration of RES. A more specific case that shows how the intermittent behavior of wind can alter the number of contingencies can be observed in Table \ref{tab:Numberofcontingency}. The number of $N-2$ contingencies can increase or decrease when compared with the case of no wind injection. One scenario that results in a lower number of $N-2$ contingencies is SC7 at $t=1400$ where the number of $N-2$ contingencies decreases from the original 71 to 67; thus making the EPS more secure under contingency conditions. A counterexample of this behavior can be observed in SC7 at $t=0$ where the number of $N-2$ contingencies increases from 71 to 84. 

\subsection{Real-time Simulation of IEEE 39 Bus System}

We further examine the effect of contingency scenarios in a real-time simulation environment. We observe the impact of intermittent wind power injections across the IEEE 39 bus system by analyzing the variability of all the bus voltages in the system. At $t=0.3s$, a $N-2$ contingency event is triggered by simultaneously disconnecting two three-phase circuit breakers. To understand the severity of losing critical elements, we disconnect the most critical pair (lines $5-8$ and $6-7$) from the $N-2$ contingency set of the IEEE 39 bus system. Fig. \ref{fig:a1} presents the $N-2$ effect that disconnecting lines $5-8$ and $6-7$ have in the test system without any wind connected. Fig. \ref{fig:a2} demonstrates the same contingency scenario (disconnection of lines $5-8$ and $6-7$) with wind power being injected to the system (Table \ref{tab:WindContingency1}: SC7 $t=800m$). In this scenario, WF1, WF2, and WF3 inject $153.70$MW, $159.20$MW, and $155.70$MW power to the test case system, respectively. An additional test scenario is run using the same contingency pairs (disconnection of lines $5-8$ and $6-7$) with different wind power injections (Table \ref{tab:WindContingency2}: SC5 $t=0m$). In this case, WF2 and WF3 inject $136.80$MW and $187.30$MW, respectively, with the results depicted in Fig.\ref{fig:a3}.

In order to understand the effect that different $N-2$ contingency pairs may have in the EPS, we perform studies using different $N-2$ pairs present in the contingency set. For these studies, we disconnect a less critical contingency pair from the $N-2$ contingency set. At $t=0.3s$, circuit breakers are tripped at lines $10-13$ and $16-21$, in a test case system without any wind power penetration, and the respective voltage variations can be observed in Fig. \ref{fig:a4}. Fig. \ref{fig:a5} depicts how the voltage variations change when wind penetration (Table \ref{tab:WindContingency1}: SC7 $t=800m$) is considered under the same contingency scenario. An additional case (Table \ref{tab:WindContingency2}: SC5 $t=0m$) with the contingency pair $10-13, 16-21$ is demonstrated in Fig. \ref{fig:a6}.

When comparing the real-time simulation results, we can observe that the most critical contingency pair (lines $5-8$ and $6-7$) causes higher voltage variations when compared to a less critical contingency pair (lines $10-13$ and $16-21$). Several buses in the EPS reach under and over-voltage values of around $0.87$ p.u. and $1.1$ p.u. for the most critical contingency pair (lines $5-8$ and $6-7$) and under and over-voltage values of around $0.92$ p.u. and $1.18$ p.u. for the less critical contingency pair (lines $10-13$ and $16-21$). Also, besides observing the voltage variations different contingency pairs can produce, we can also observe, in some cases, how the intermittent behavior of wind power helps to mitigate the severity of line overloads. Figs. \ref{fig:a4} -- \ref{fig:a6} demonstrate this behavior. For instance, in Fig. \ref{fig:a5}, most buses of the power system have voltage measurements that are closer to the nominal $1.0$ p.u value. On the other hand, the case in Fig. \ref{fig:a6} shows the opposite, since some voltage values measured at some buses are farther apart from $1.0$ p.u when compared with the case where no wind power injection is included, i.e., Fig. \ref{fig:a4}. Our results demonstrate how important is to coordinate the amount of wind power as it penetrates the system. For example, the authors in \cite{bai2016robust} proposed a scheme for power systems to maintain $N-1$ security within different levels of wind power injection. In addition, a dynamic reserve allocation of DFIG wind farms is presented in \cite{chang2008dynamic} to sustain system frequency stability. In our case, the results not only demonstrate the variation of $N-2$ contingency numbers but also how these results can be used to control the penetration level of wind farms to increase the $N-2$ secure operational range of power systems.

\section{Results: The Effectiveness of the Proposed Cybersecurity Assessment}
\label{s:result2}

This section presents our experimental results that demonstrate the effectiveness of the proposed cybersecurity assessment approach. We evaluate the efficacy of the process according to the optimal attack transition policies given as outputs. In this part, we provide the experimental setup for the presented test cases, the D$Q$N agent model implementation details, and its corresponding hyperparameters. Six test case systems are used to demonstrate the number of transitions needed to identify the optimal attack path for the corresponding case. Furthermore, the performance of the D$Q$N model is evaluated according to the obtained rewards and losses, i.e., convergence for each test case. Finally, the effectiveness of the D$Q$N, used to solve the transition model, is verified by comparing it to other transition-path policy-finding methods, and specifically to the: \textit{(i)} \emph{random policy search}, \textit{(ii)} \emph{depth-first search (DFS)}, \textit{(iii)} \emph{Dijkstra's shortest path} algorithm, and an \textit{(iv)} \emph{IVSS-based D$Q$N} model. 

\begin{table}[]
\centering
\caption{D$Q$N hyperparameters.}
\label{table:dqnhyperparameters}
\begin{tabular}{||c|c||}
\hline\hline
\multicolumn{2}{||c||}{\textbf{DQN Hyperparameters}} \\ \hline\hline
Num. of hidden layers & System size  \\ \hline
Num. of hidden neurons & 1000\\ \hline
Learning rate ($\alpha$) & 0.005 \\ \hline
Activation fcn. & ReLu \\ \hline
Optimizer & Adam \\ \hline
Max. steps & 1000 \\ \hline
Max. episodes & 100 \\ \hline
Replay memory buffer (samples) & 1000 \\ \hline
Exploration policy & \begin{tabular}[c]{@{}c@{}}$\epsilon$-greedy \\ linear decay\end{tabular} \\ \hline
Exploration rate ($\epsilon$) & 0.01 \\ \hline\hline
\end{tabular}
\end{table}

\subsection{Experimental Setup and DQN Hyperparameters}

The RL D$Q$N model is trained and tested on a 64-bit machine with an Intel Core i7-7600U, 2.8GHz, and 16.00GB of memory. The proposed algorithm is implemented in \texttt{Julia}, a high-level, high-performance, dynamic programming language. The D$Q$N solver for POMDP is provided in \cite{hsinimplementation}. The source files and models associated with this work can be found at \cite{github_dss}. The D$Q$N hyperparameters are presented in Table \ref{table:dqnhyperparameters}.

\begin{table}[t]
\tabcolsep=0.03cm
\caption{Average number of transitions and timing for cyberphysical attacks.}
\renewcommand{\arraystretch}{1.2}
\begin{center}
\begin{tabular}{||c|c|c|c|c|c||}
\hline \hline

\multirow{2}{*}{\centering {Case Name}} &{\centering {$\#$}}&{$\#$}&{\textbf{$\#$}}&{$T_{Tr}$}&{$T_{To}$} \\ 
&{\centering {{Trans}}}&{$PV$} &{$PQ$}&($sec$) &($sec$)\\\hline

{IEEE 30 Bus System} & {$9.8$} & 2.6& 9.8&{76.2} &{83.6} \\ \cline{1-6}

{IEEE 39 Bus System} & {$8.8$} &0.4 &4.6 &{89.3}&{96.9}\\ \cline{1-6}

\hspace{0.05in} {IEEE 39 Bus System +  Wind} \hspace{0.05in} & \multirow{2}{*}{$11.6$} &\multirow{2}{*}{$0.6$}&\multirow{2}{*}{5.8} &\multirow{2}{*}{90.1}&\multirow{2}{*}{97.6}\\

W1 (Table \ref{tab:WindContingency1}: SC7, $t=800m$)&&&&& \\\cline{1-6} 

\hspace{0.05in} {IEEE 39 Bus System + Wind} \hspace{0.05in} & \multirow{2}{*}{$14.8$} &\multirow{2}{*}{$2.8$}&\multirow{2}{*}{7.6} &\multirow{2}{*}{87.8}&\multirow{2}{*}{95.5}\\

W2  (Table \ref{tab:WindContingency2}: SC5, $t=0m$)&&&&& \\\cline{1-6} 

{UIUC 150 Bus System} & {$34.0$} &2.4 &28.4&{237.9}& {245.6}\\ \cline{1-6}

{Polish 2383 Bus System} & {$35.0$} &5.4 &30.8&{1459.5}& {1506.7}\\ \cline{1-6}

\hline \hline
\end{tabular}
\label{tab:ComparisionTable}
\end{center}
\end{table}

\subsection{Cybersecurity Assessment: Attack-Path Transition Results}

In order to demonstrate the efficacy of the proposed cybersecurity assessment process, we use six test case power systems related with the contingency studies in Section \ref{s:result} (Table \ref{tab:Numberofcontingency}): \textit{(a)} IEEE 30 bus system, \textit{(b)} IEEE 39 bus system, \textit{(c)} IEEE 39 bus system with wind W1 (Table  \ref{tab:WindContingency1} SC7 at $t=800m$), \textit{(d)} IEEE 39 bus system with wind W2 (Table  \ref{tab:WindContingency2} SC5 at $t=0m$), \textit{(e)} UIUC 150 bus system, and the \textit{(f)} Polish 2383 bus system. Based on the identified critical $N-2$ pairs, the malicious agent begins at a random initial state and finds the optimal attack-path transition policy to the existing and most critical $N-2$ contingencies. A contingency is identified when one of the two buses has been visited by the agent. In Table \ref{tab:ComparisionTable}, we show the number of transitions required to reach both critical contingencies as well as the number of $PV$ and $PQ$ buses visited by the agent. For each comparison, five random initial states are selected for each test system, and the average results are presented. For example, the IEEE 39 bus system requires an average of $8.8$ transitions to correctly identify the most critical contingency pair. 
During the transitions, an average number of $0.4$ generation ($PV$) and $4.6$ load ($PQ$) buses need to be visited, i.e., compromised, by the agent. $T_{Tr}$ is the training and evaluation time (in seconds) needed for the D$Q$N to `learn' the optimal attack path for different cases, and $T_{To}$ is the total time (in seconds) required to complete the process. The utilization of the Polish 2383 bus system in our experimental results aids in the evaluation of our proposed process with a realistic large-scale EPS. As seen in Table \ref{tab:ComparisionTable}, the training and evaluation process of the D$Q$N in a typical computer with 2.8GHz CPU and 16.00GB RAM requires an average of 1459.5 seconds (approximately 24 minutes), and a total running time of 1506.7 seconds (approximately 25 minutes). Note that the running time of the entire process could be reduced by decreasing the number of hidden neurons in the D$Q$N model. The results demonstrate that the proposed cybersecurity process can be used in tandem with medium and long-term control and planning applications. On the other hand, the proposed approach would require high computing power in order to be integrated into very short-term decision making processes \cite{timescales}.

\begin{figure}[t]
\centerline{\includegraphics[width=0.85\linewidth]{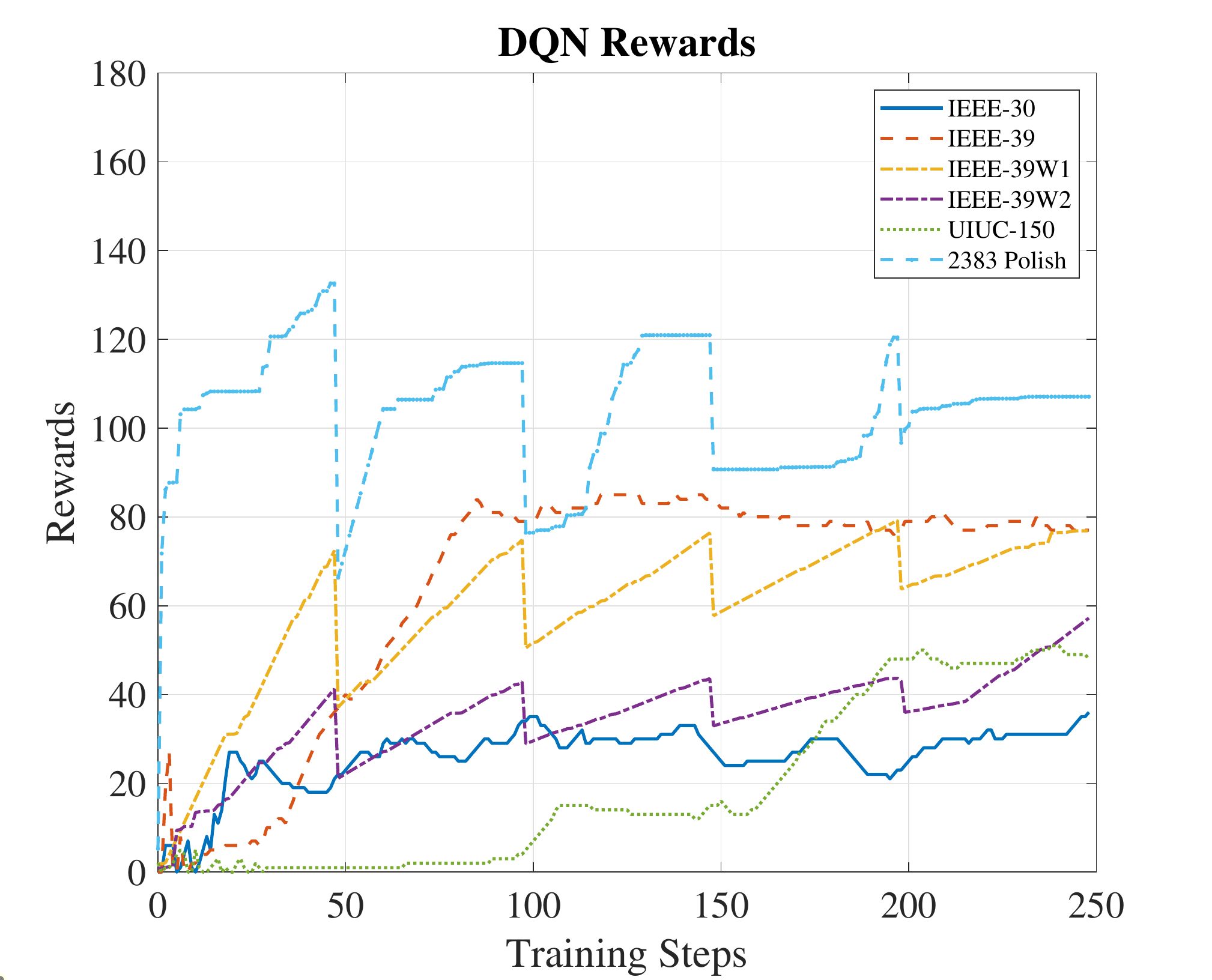}}
\caption{D$Q$N rewards for bus test systems.} 
\label{fig:reward}
\end{figure}

\begin{figure}[t]
\centerline{\includegraphics[width=0.85\linewidth]{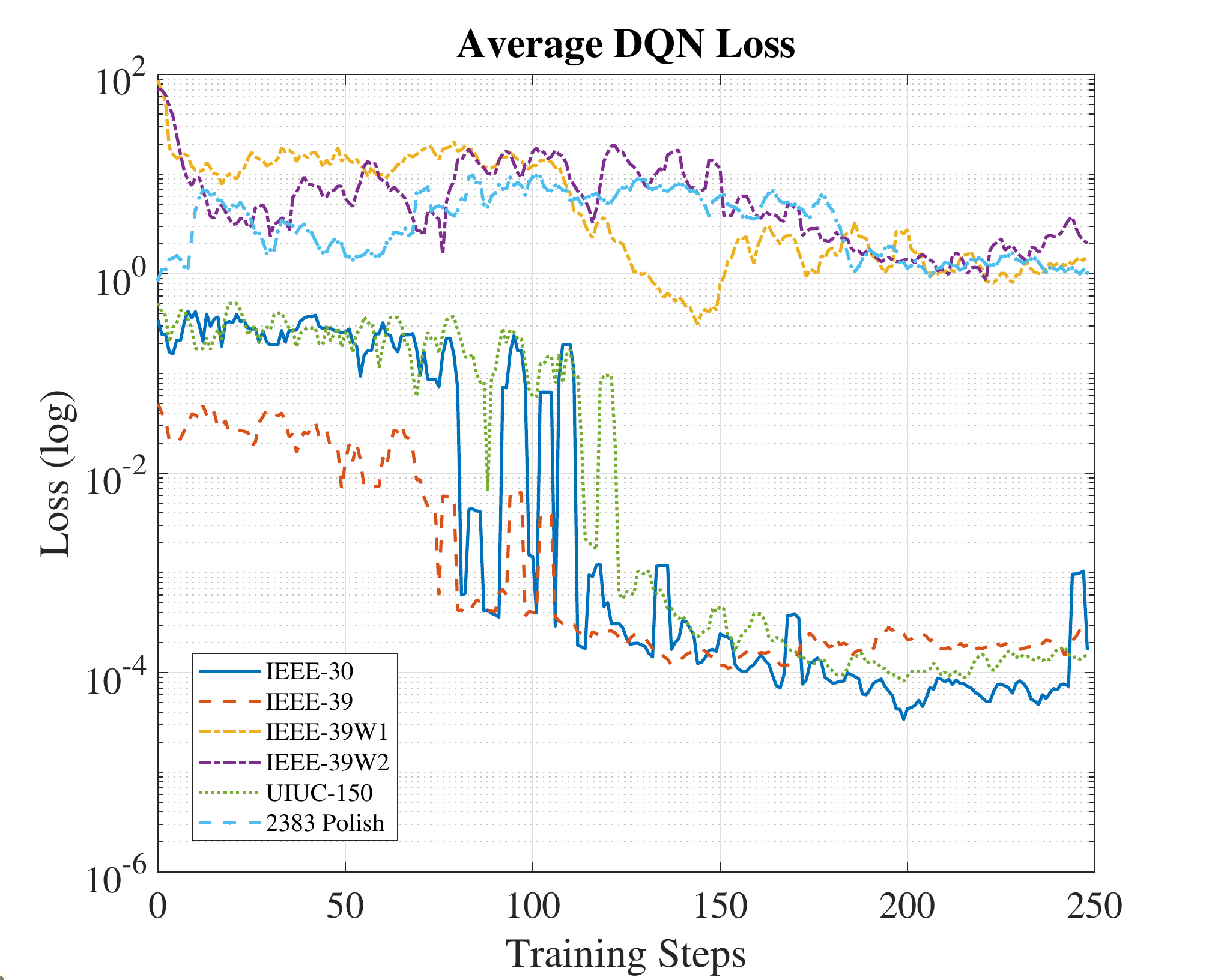}}
\caption{Average D$Q$N loss for bus test systems.} 
\label{fig:loss}
\end{figure}

\begin{figure*} \centering    
\subfigure[] { \label{fig:5}     
\includegraphics[width=5.72cm]{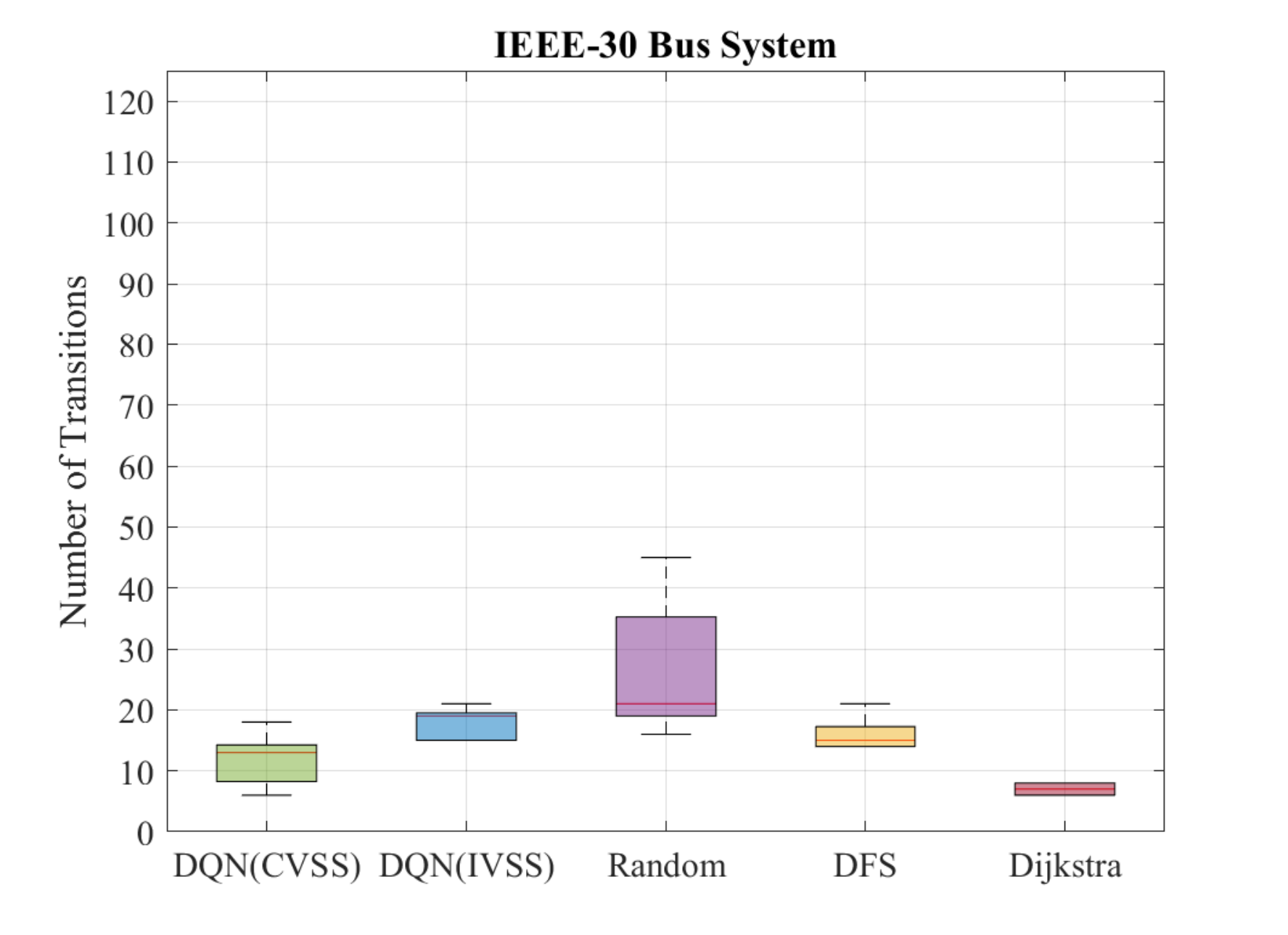}  
 }
\subfigure[] { \label{fig:6}     
\includegraphics[width=5.7cm]{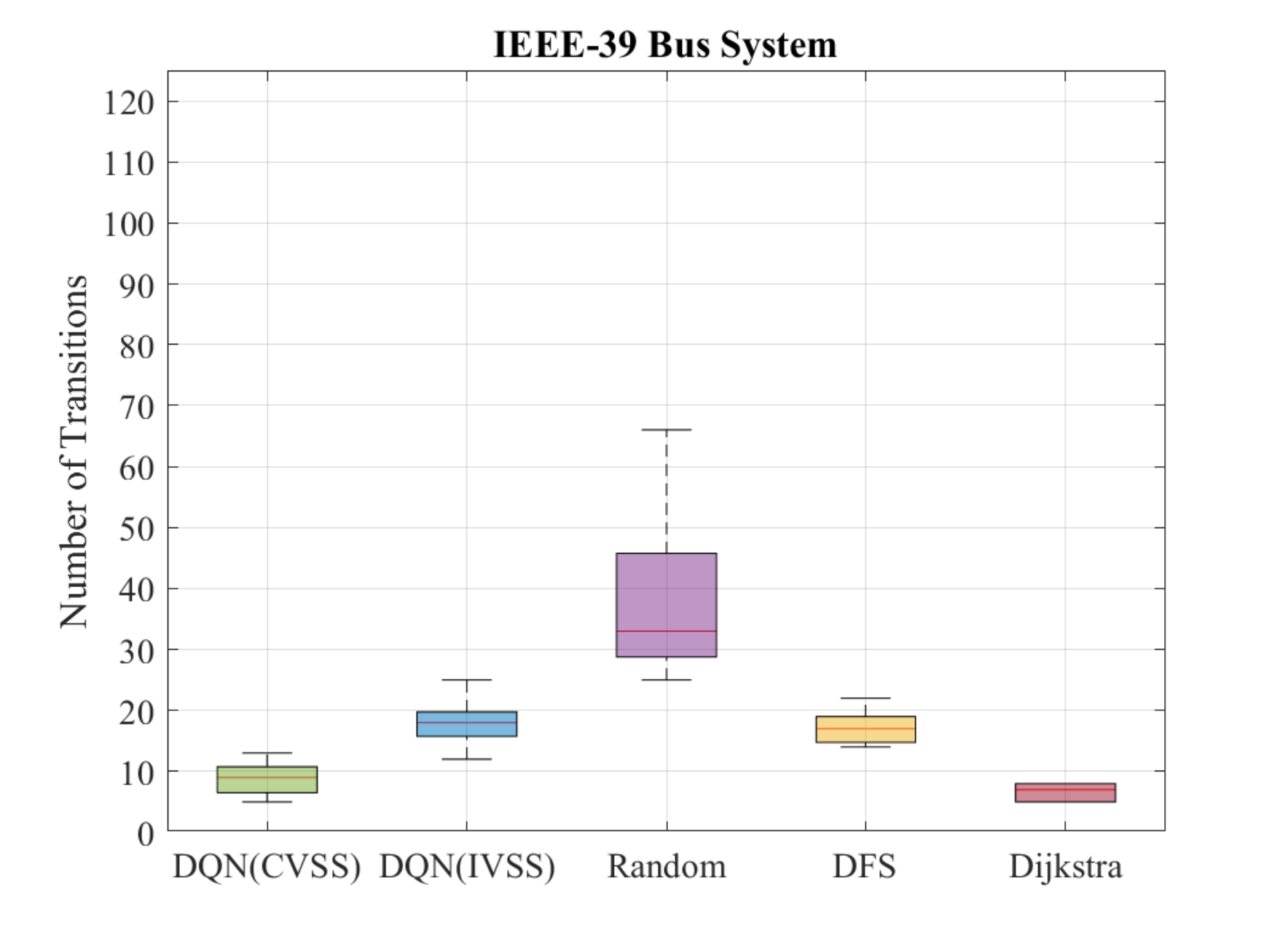}}
\subfigure[] { \label{fig:7}     
\includegraphics[width=5.7cm]{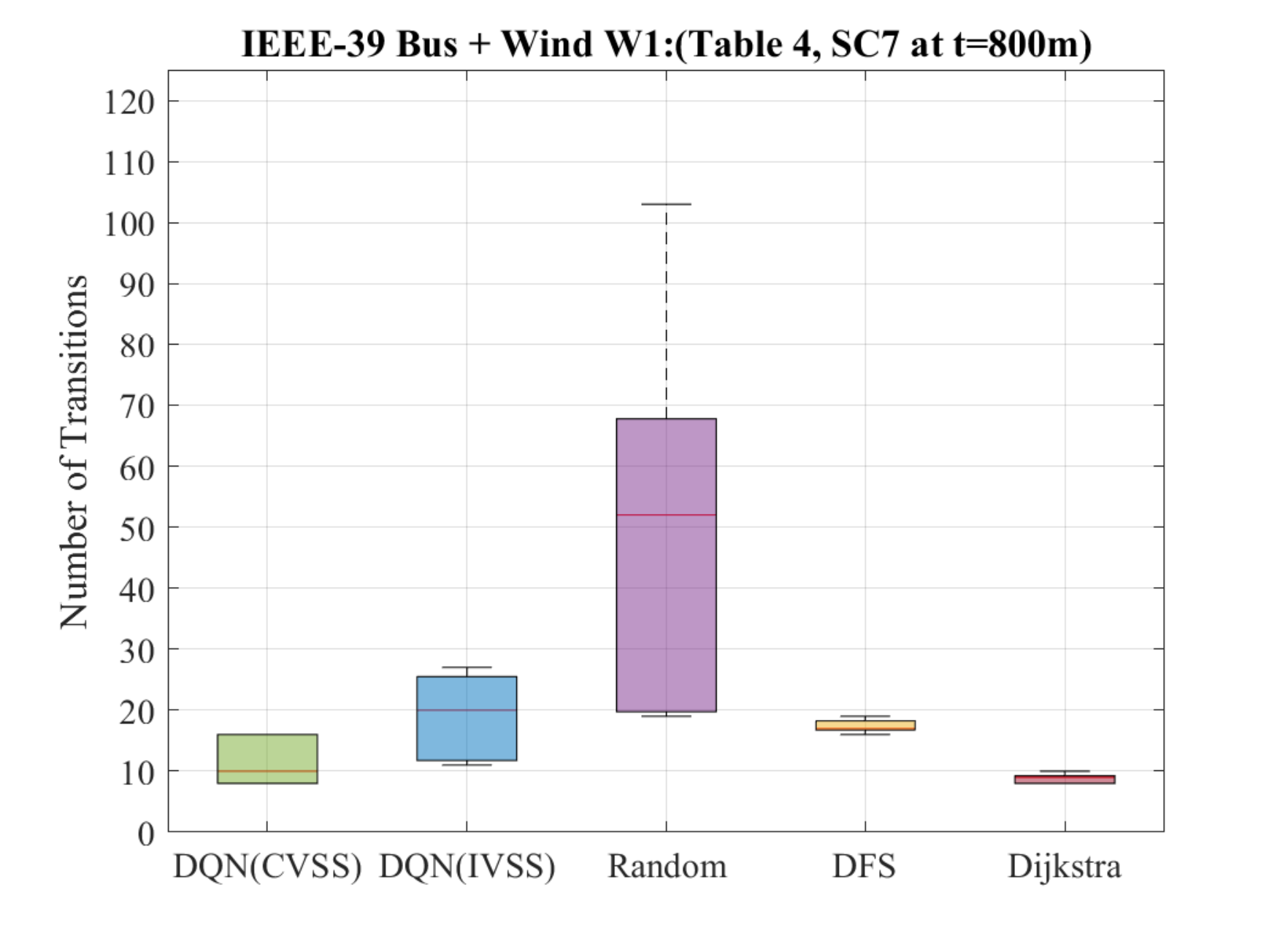}}
\\
\subfigure[] { \label{fig:8}     
\includegraphics[width=5.7cm]{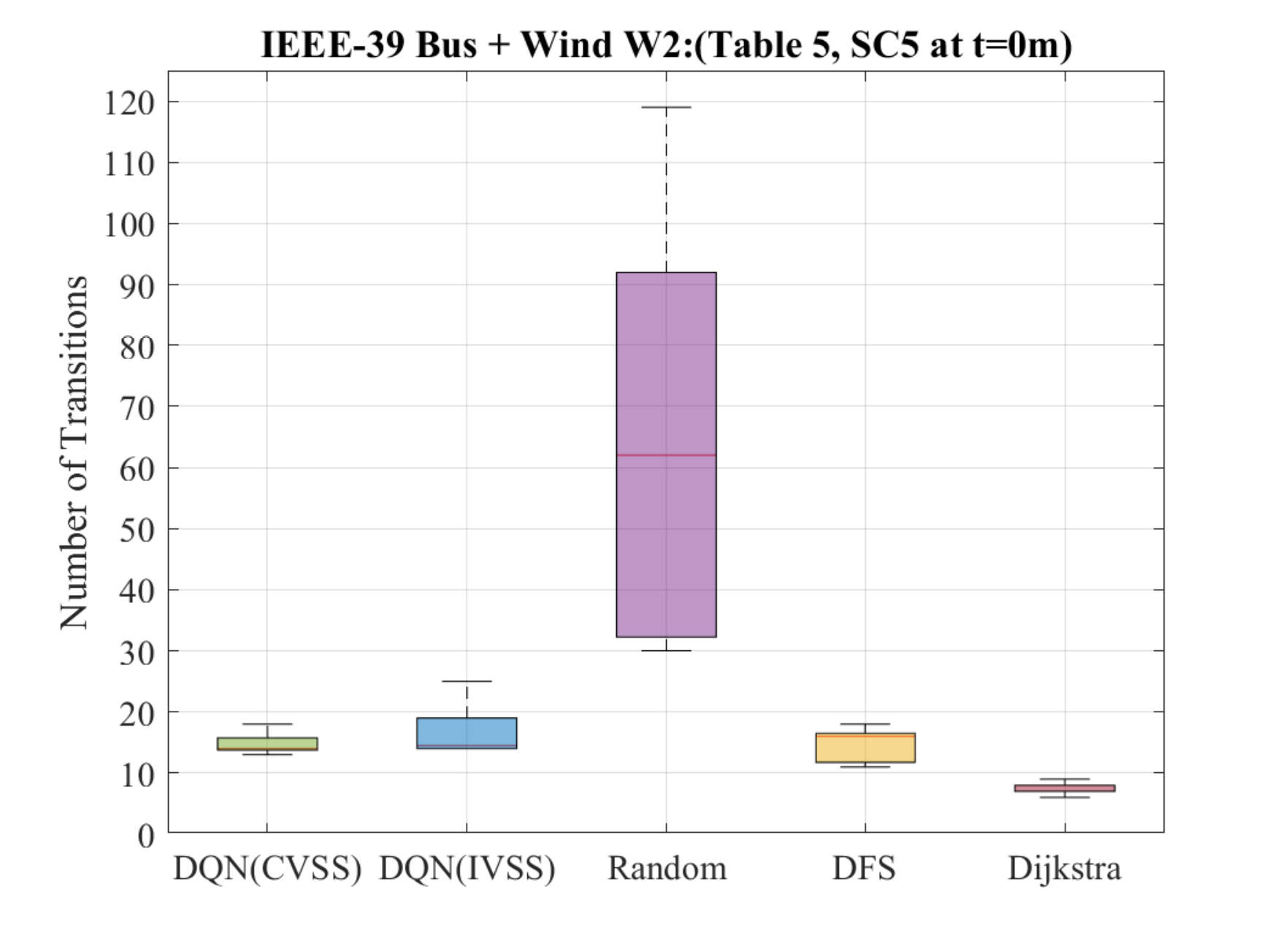}}
\subfigure[] { \label{fig:9}     
\includegraphics[width=5.7cm]{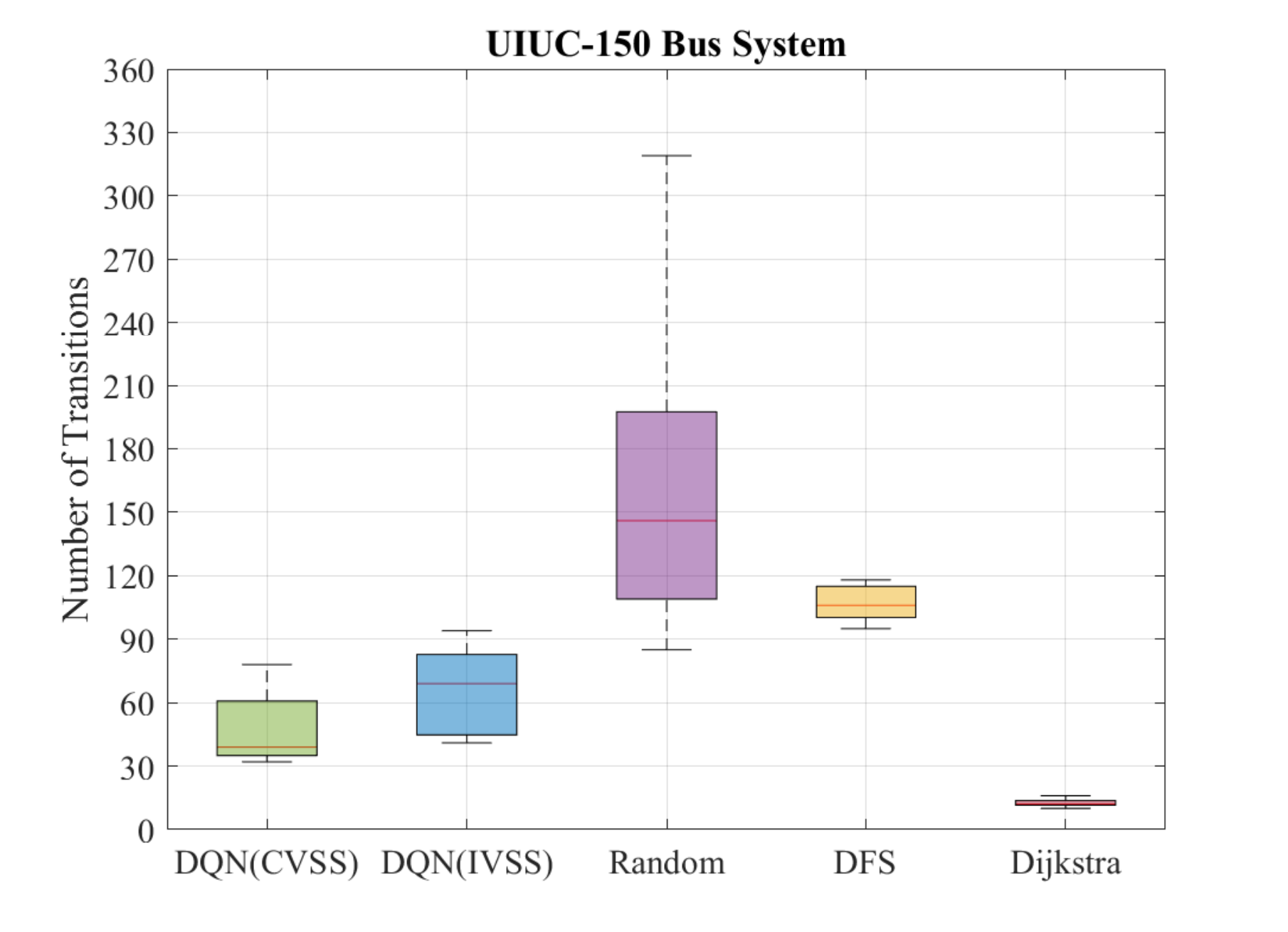}}
\subfigure[] { \label{fig:10}     
\includegraphics[width=5.7cm]{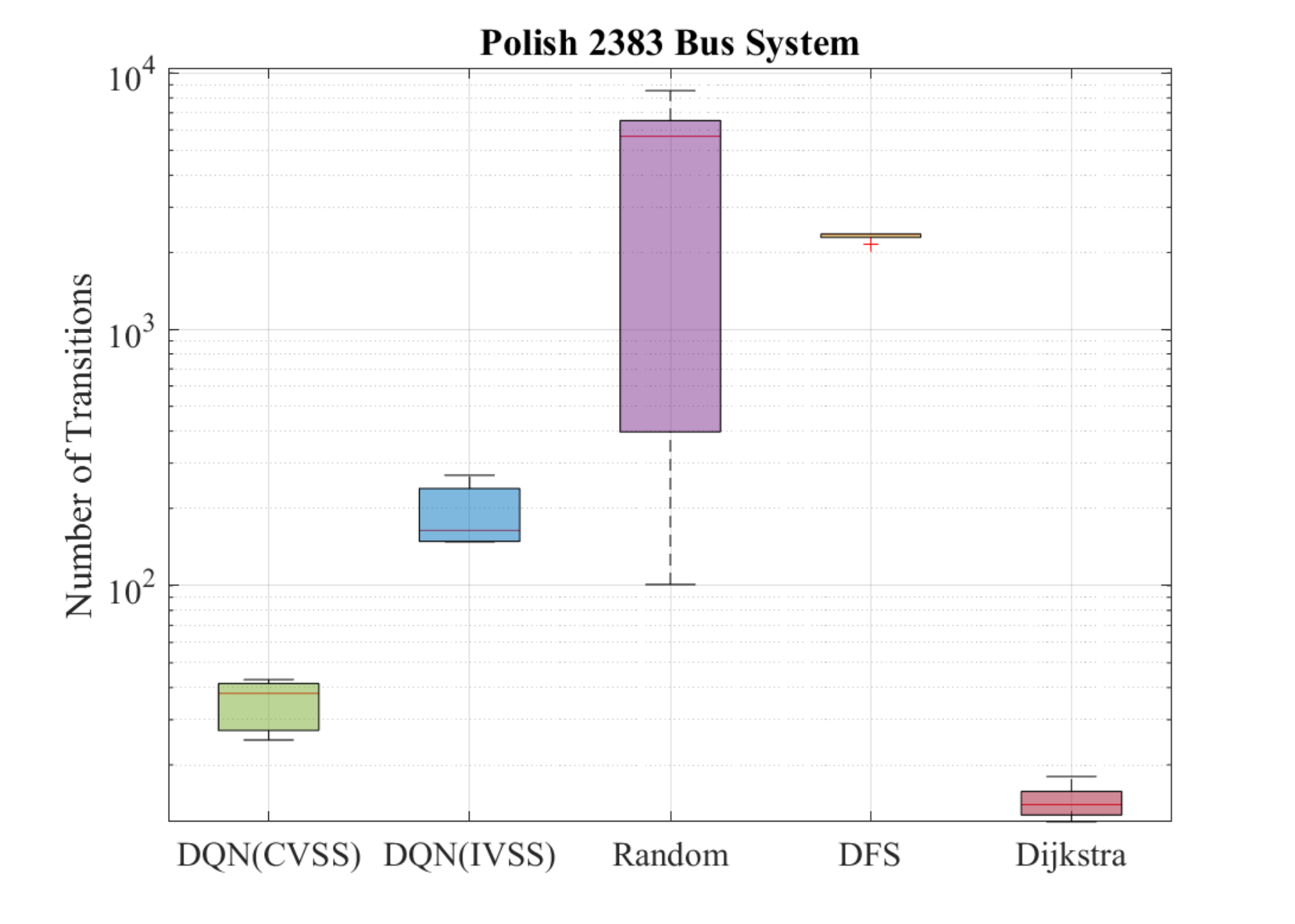} }

\caption{Number of transitions needed for \emph{D$Q$N (CVSS-based)}, \emph{D$Q$N (IVSS-based)}, \emph{DFS}, \emph{random}, and \emph{Dijkstra} transition policies for \subref{fig:5} IEEE 30 bus system, \subref{fig:6} IEEE 39 bus system, \subref{fig:7} IEEE 39 bus system + Wind W1 (Table  \ref{tab:WindContingency1}: SC7, $t=800m$), \subref{fig:8} IEEE 39 bus system + Wind W2 (Table \ref{tab:WindContingency2}: SC5, $t=0m$), \subref{fig:9} UIUC 150 bus system, and the \subref{fig:10} Polish 2383 bus system.} 
\label{fig:resultsattacks}  
\end{figure*}

\subsection{DQN Rewards and Loss Convergence}

As mentioned in Section \ref{s:method}, the D$Q$N aims to minimize the loss between the target value and the predicted value. The D$Q$N agent learns the optimal policy as this loss is minimized. Here, we verify and evaluate the performance of our proposed approach by examining the convergence of the D$Q$N loss during the training process. We also show how the average reward gradually increases at each step, for each test case, up to $250$ training steps. It should be noted that the total number of training steps used is $500$ while the update frequency of the plot is set to $2$, thus only $250$ steps can be observed in the graph. Fig. \ref{fig:reward} shows the rewards for each test case system and Fig. \ref{fig:loss} shows the corresponding loss for each case. As shown in Fig. \ref{fig:reward}, the D$Q$N agent progressively `learns' how to maximize the cumulative rewards in each test case system. At the same time, as the agent `learns', the loss keeps decreasing until it converges to a minimum value as depicted in Fig. \ref{fig:loss}. These results showcase the training process of the D$Q$N agent and its performance on all bus test case systems.

\begin{figure*} \centering    
\subfigure[] { \label{fig:10a}     
\includegraphics[width=7.7cm]{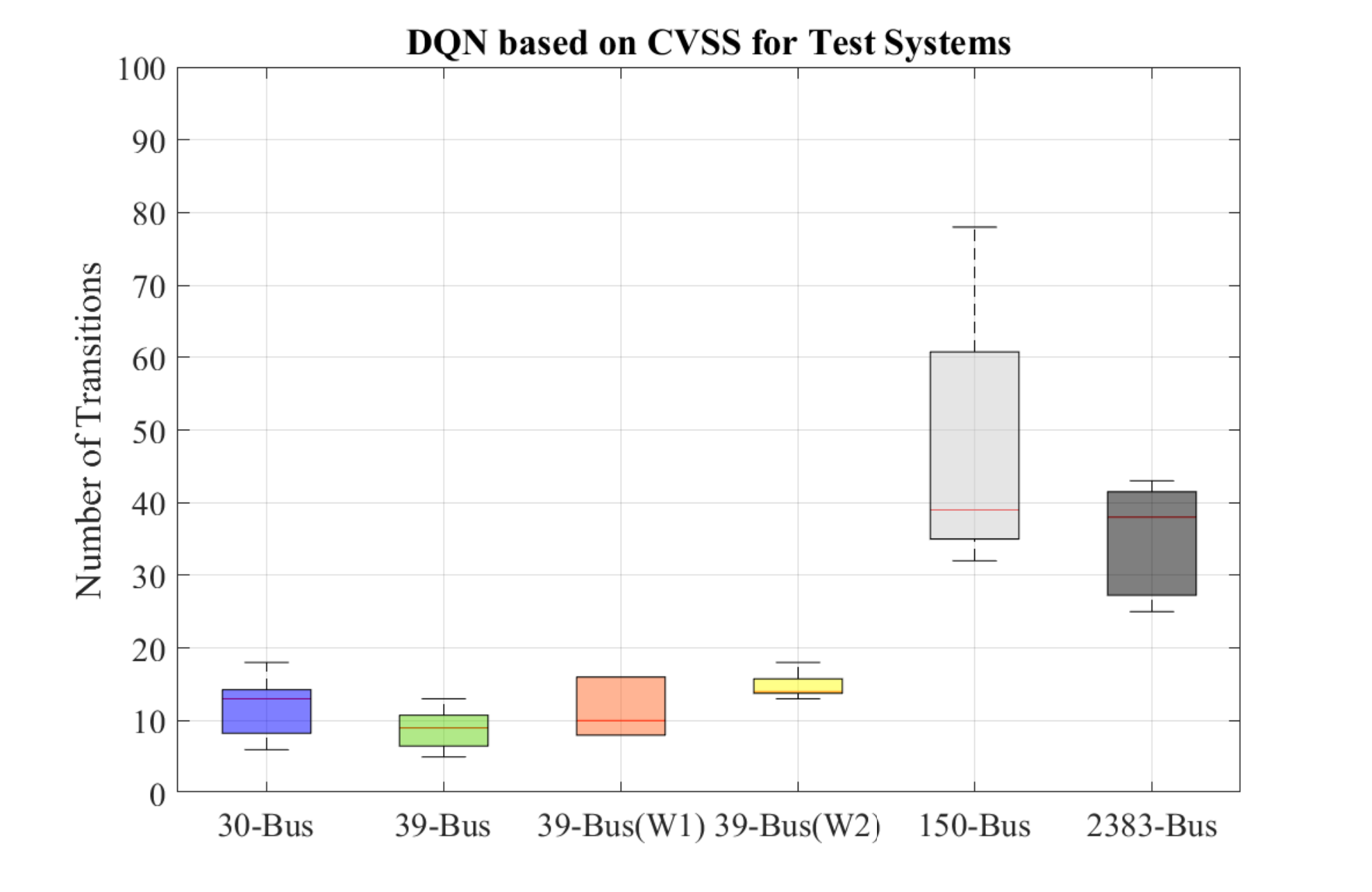}  
 }
\subfigure[] { \label{fig:10b}     
\includegraphics[width=7.7cm]{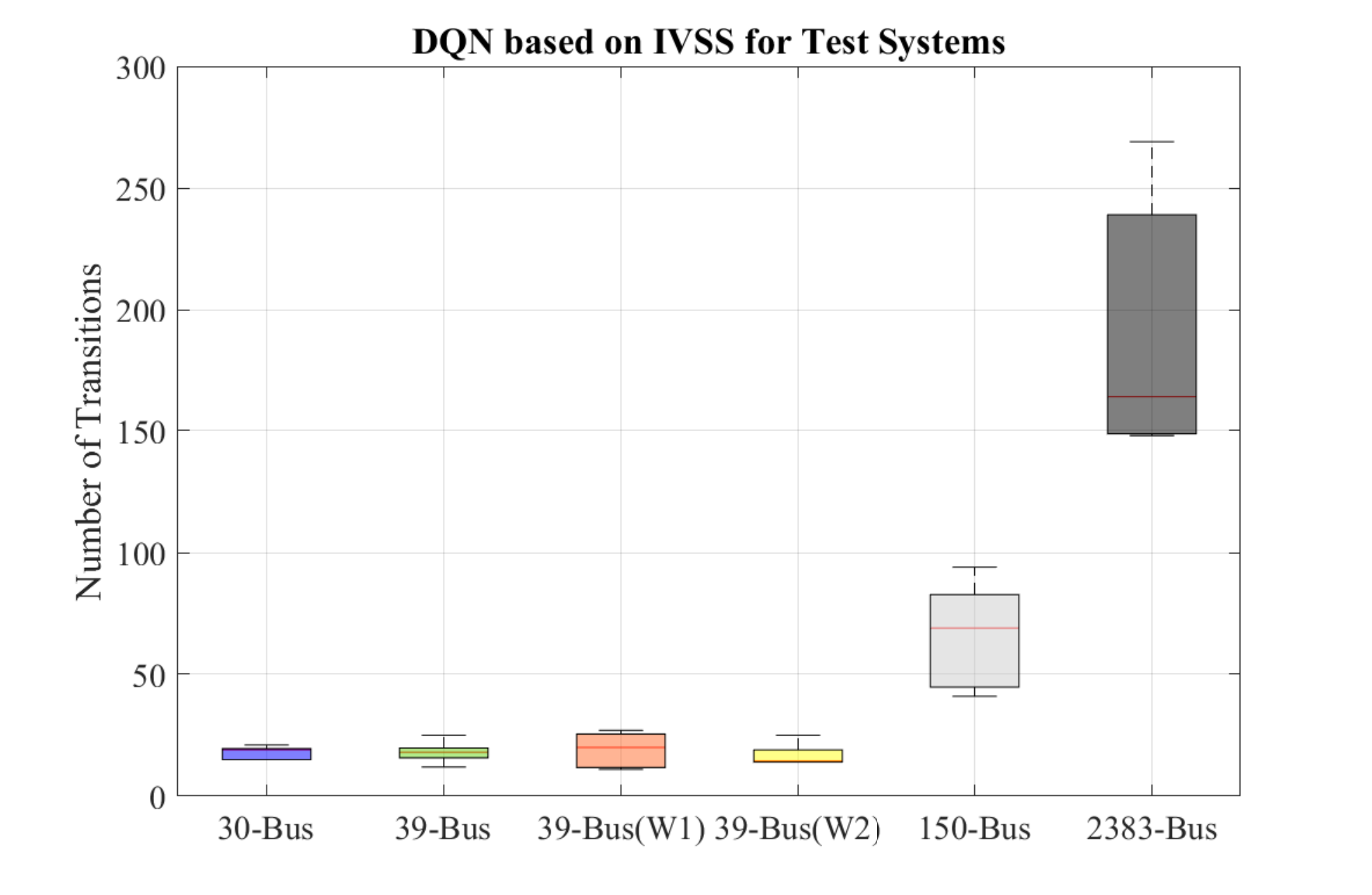}}

\caption{ Number of transitions needed for \emph{D$Q$N (CVSS-based)} and  \emph{D$Q$N (IVSS-based)} on all test systems evaluated. These figures demonstrate the scalability of the proposed approach as the number of buses increases.}
\vspace{-2mm}
\label{fig:resultscalability}  
\end{figure*}

\subsection{Effectiveness of DQN: Comparison with other Transition Techniques}

The effectiveness of using a D$Q$N model in our cybersecurity assessment process is demonstrated by comparing our D$Q$N agent based on the CVSS scoring system with different techniques that could be used to find the optimal attack transition policy in a graph. The techniques used to compare the performance of the proposed D$Q$N are: \textit{(i)} \emph{random policy search}, \textit{(ii)} \textit{DFS}, \textit{(iii)} \emph{Dijkstra's shortest path} algorithm, and \textit{(iv)} \emph{IVSS-based D$Q$N} model. The \emph{random} transition technique provides a baseline, or naive case, where transitions are performed randomly, i.e., without any intelligent control mechanisms. \emph{DFS} is a searching technique for traversing a tree structure by starting from an arbitrary root node and exploring each branch as far as possible before going back to the root node and continuing to the next branch. \emph{Dijkstra's} algorithm is a more sophisticated way of finding an optimal path through a graph structure.  \emph{Dijkstra's} algorithm is used to solve shortest-path problems in non-negative weighted graphs by finding an acyclic path between a source and a target node with the minimum transition cost. Both \emph{DFS} and \emph{Dijkstra's} search policies need full observability of the network, hence, for testing purposes in those two cases, we assume full observability of the system and its corresponding contingency pair. Finally, the \emph{IVSS-based} D$Q$N model is designed to evaluate the differences between the CVSS and IVSS vulnerability assessment criteria. 

The tests conducted are run using the power system test cases presented in Table \ref{tab:ComparisionTable}. For each case, five random initial states are selected and the average number of transitions is calculated. The maximum, minimum, and average number of transitions for each case are shown in the box plots presented in Fig.\ref{fig:resultsattacks}. From Figs. \ref{fig:5} -- \ref{fig:10}, we can observe that, in general, the results of the D$Q$Ns-based transition techniques tend to require fewer number of transitions, i.e., are more efficient, when compared with the \emph{random} and the \textit{DFS} transition techniques. When compared with \emph{Dijkstra's} algorithm, our D$Q$N implementation performs slightly worse due to its iterative learning process. However,  \emph{Dijkstra's shortest path} algorithm has the major disadvantage of requiring full system observability. The results demonstrate the advantages of using D$Q$N as the main solver technique for our proposed process. Finally, it can also be observed from Fig. \ref{fig:resultsattacks} that using CVSS v.3.1 has major advantages when compared to the IVSS scoring system. The CVSS-based D$Q$N consistently requires fewer number of transitions in all evaluated test cases.

To understand the scalability of the proposed D$Q$N approach based on CVSS and IVSS, the number of transitions for the different test systems evaluated are plotted in the box plots shown in Figs. \ref{fig:10a} and \ref{fig:10b}. In these figures, we can observe that as the number of buses increase, the number of transitions also increases but not in an exponential fashion. Additionally, the results depicted in the figures demonstrate that in almost all test systems the D$Q$N based on CVSS requires a smaller number of transitions than the D$Q$N approach based on IVSS. An example of the improved performance when using the CVSS-based approach can be observed when comparing the number of transitions required for the Polish 2383 test system.

\section{Conclusions and future work}\label{s:conclusion}
In this paper, we present a cybersecurity assessment approach designed to assess the cyberphysical security of EPS with high penetration of wind. The proposed process assumes that adversaries could leverage OSINT to perform contingency analysis. Based on the contingency results and identified exploitable cyberphysical vulnerabilities via an adapted CVSS metric, an optimal attack transition policy is generated that can be potentially leveraged to cause major outages in an EPS. The results provided by the proposed process are also critical for improving cybersecurity visibility for system operators and stakeholders; it provides information regarding the most critical attack-path an adversary must follow to severely compromise the system alongside with information about the most vulnerable elements in the EPS at a particular time. The proposed approach is tested using real-time simulation, realistic data from various actual wind energy systems, and various test case power systems. Additionally, results regarding the training and convergence of the D$Q$N agent, proposed as the main optimal attack-path transition technique, are presented and compared with other competing techniques. These results demonstrate the applicability of the cybersecurity assessment approach in modern EPS.

The limitations of the cybersecurity assessment approach presented in this work include mostly the assumptions related to the threat model: \textit{(i)} The contingency analysis can only be performed when the attacker has sufficient power system data acquired using OSINT techniques. Without the necessary information, the set of contingencies cannot be correctly identified. \textit{(ii)} The cyber system network graph is assumed to be isomorphic with the physical system graph, indicating that the topology of the communication network is mapped one-to-one with the topology of the physical system. This assumption may not be necessarily true on some systems, since the physical and communication networks could have different network topologies.

Based on the limitations discussed, future work will focus on: \textit{(i)} Exploring potential defense strategies based on moving target defense methods that could be used to enhance the overall system security and resilience by dynamically updating time-varying parameters within the control system of EPS (act as a moving target), thus, limiting adversaries understanding of the cyberphysical EPS model. \textit{(ii)} Analyzing and investigating other DRL solvers that can be adapted into the proposed cybersecurity assessment approach (e.g., UCB, A3C, or TRPO). \textit{(iii)} Investigating how transitions, in the proposed cybersecurity assessment approach, are affected in scenarios where the cyber and physical networks are not assumed isomorphic; examined by using real-time co-simulation testbeds.

\bibliographystyle{IEEEtran} 
\bibliography{biblio}
\section*{} \label{s:bio}
\vskip -2\baselineskip plus -1fil

\begin{IEEEbiography}[{\includegraphics[width=1in,height=1.25in,clip,keepaspectratio]{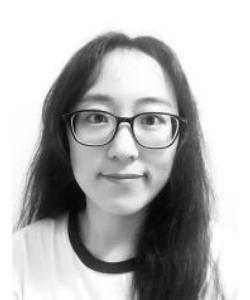}}]{XiaoRui Liu} (S'20) received her M.S. degree in Electrical Engineering from the Florida State University, Tallahassee, FL, USA, in 2017. She is currently pursuing the Ph.D. degree in Electrical Engineering at Florida State University, Tallahassee, FL, USA. Her research interest includes real-time simulation of power systems, cybersecurity, and machine learning. 

\end{IEEEbiography}
\vskip -2\baselineskip plus -1fil
\begin{IEEEbiography}[{\includegraphics[width=1in,height=1.25in,clip,keepaspectratio]{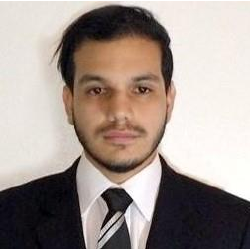}}]{Juan Ospina}~(S'13-M'20) is a Postdoctoral Research Associate with Florida State University and the Center for Advanced Power Systems, Tallahassee, Florida. He received a dual B.Sc. degree in Electrical and Computer Engineering in 2016, an M.S. in Electrical Engineering in 2018, and a Ph.D. in Electrical Engineering in 2019 from Florida State University, Tallahassee, Florida, USA. He is an IEEE and IEEE PES member. His research interests include the development of intelligent systems for electric power systems (EPS) and smart-grid applications, machine learning and reinforcement learning models for DER control, renewable energy integration, cybersecurity, and real-time simulation.
\end{IEEEbiography}

\vskip -2\baselineskip plus -1fil
\begin{IEEEbiography}[{\includegraphics[width=1in,height=1.25in,clip,keepaspectratio]{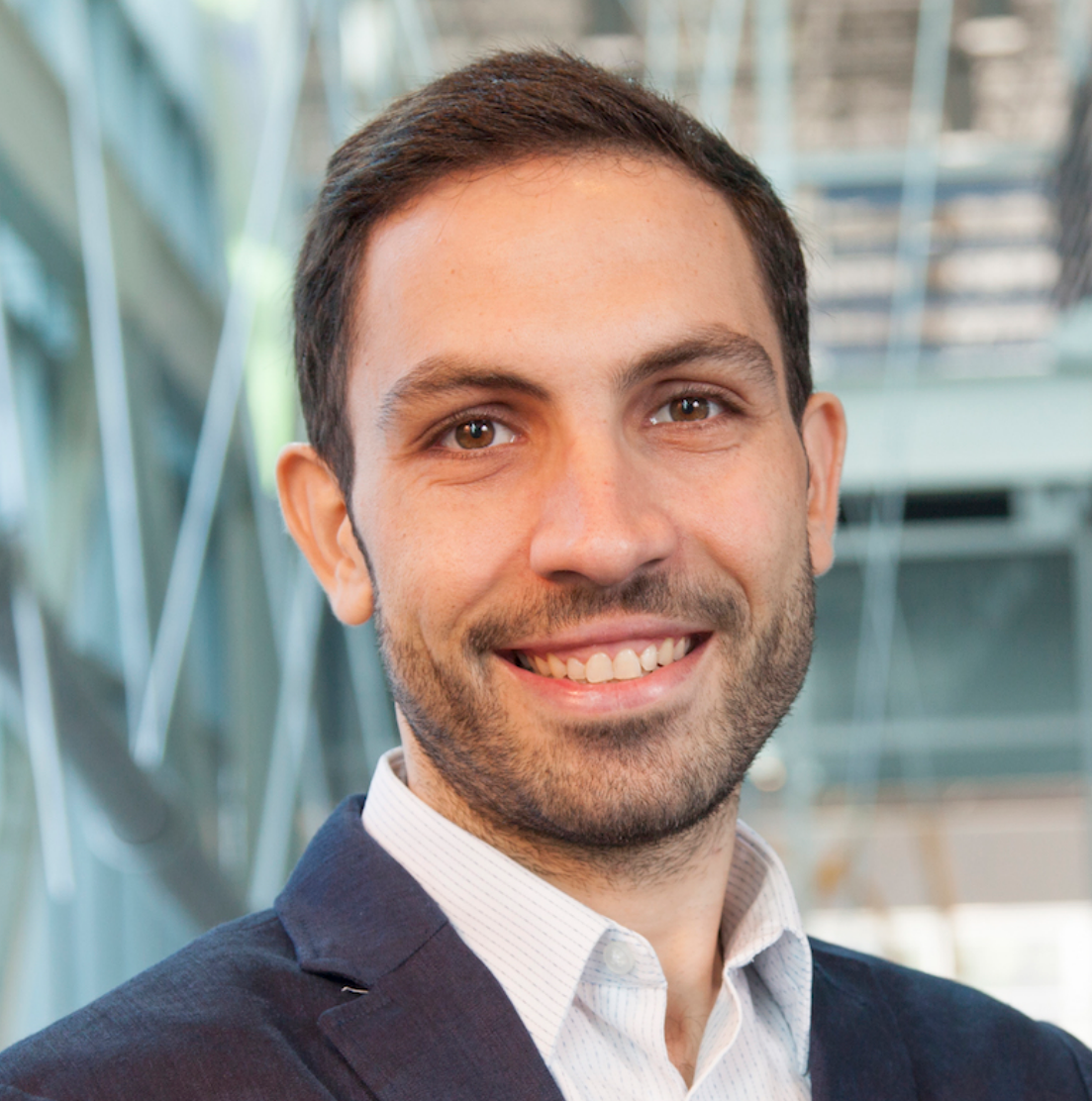}}]{Charalambos Konstantinou}~(S'11-M'18-SM'20) is an Assistant Professor of Electrical and Computer Engineering with Florida A\&M University and Florida State University (FAMU-FSU) College of Engineering and the Center for Advanced Power Systems, Florida State University, Tallahassee, FL. He received a Ph.D. in Electrical Engineering from New York University, NY, in 2018. His research interests include cyberphysical and embedded systems security with focus on power systems. He is the recipient of the 2020 Myron Zucker Student-Faculty Grant Award from IEEE Foundation, the Southeastern Center for Electrical Engineering Education (SCEEE) Young Faculty Development Award 2019, and the best paper award at the International Conference on Very Large Scale Integration (VLSI-SoC) 2018. 
\end{IEEEbiography}

\EOD

\end{document}